\documentclass[
 reprint,
superscriptaddress,
 amsmath,amssymb,
 aps,
 prl,
 longbibliography
]{revtex4-2}

\usepackage[version=4]{mhchem}
\usepackage{xcolor}
\usepackage{graphicx}
\usepackage{dcolumn}
\usepackage{bm}
\usepackage{mathtools}
\usepackage{siunitx}
\usepackage{verbatim}
\usepackage{titletoc}

\makeatletter
\def\maketitle{%
  \@author@finish
  \title@column\titleblock@produce
  \suppressfloats[t]%
}
\makeatother

\usepackage{verbatim}

\makeatletter
\renewcommand\email[1]{%
  \thanks{Contact author: \href{mailto:#1}{#1}}%
}
\makeatother

\begin{document}

\preprint{APS/123-QED}

\def\mytitle{False Metallization in Short-Ranged Machine Learned Interatomic Potentials}
\title{\mytitle}

\author{Isaac J. Parker}
\affiliation{Engineering Laboratory, University of Cambridge, Trumpington St., Cambridge, UK}
\affiliation{Cavendish Laboratory, University of Cambridge, J. J. Thomson Ave., Cambridge, UK}
\affiliation{Lennard-Jones Centre, University of Cambridge, Trinity Ln, Cambridge, CB2 1TN, UK}

\author{Mandy J. Hoffmann}
\affiliation{Engineering Laboratory, University of Cambridge, Trumpington St., Cambridge, UK}
\affiliation{Yusuf Hamied Department of Chemistry, University of Cambridge, Lensfield Road, Cambridge, UK}
\affiliation{Lennard-Jones Centre, University of Cambridge, Trinity Ln, Cambridge, CB2 1TN, UK}

\author{William J. Baldwin}
\email{wjb48@cam.ac.uk}
\affiliation{Engineering Laboratory, University of Cambridge, Trumpington St., Cambridge, UK}
\affiliation{Lennard-Jones Centre, University of Cambridge, Trinity Ln, Cambridge, CB2 1TN, UK}


\author{Shuang Han}%
\affiliation{%
BASF SE, Group Research, Carl-Bosch-Straße 38, 67056 Ludwigshafen, Germany
}%
\author{Srishti Gupta}%
\affiliation{%
BASF Chemicals India Pvt. Ltd., Plot No.12, TTC Area, Thane Belapur Road, Turbhe, Navi Mumbai 400705, India
}%

\author{Kara D. Fong}
\affiliation{Division of Chemistry and Chemical Engineering, California Institute of Technology, Pasadena, California 91125, USA}
\affiliation{Marcus Center for Theoretical Chemistry, California Institute of Technology, Pasadena, California 91125, USA}

\author{Angelos Michaelides}
\affiliation{Yusuf Hamied Department of Chemistry, University of Cambridge, Lensfield Road, Cambridge, UK}
\affiliation{Lennard-Jones Centre, University of Cambridge, Trinity Ln, Cambridge, CB2 1TN, UK}

\author{Christoph Schran}
\email{cs2121@cam.ac.uk}
\affiliation{Cavendish Laboratory, University of Cambridge, J. J. Thomson Ave., Cambridge, UK}%
\affiliation{Lennard-Jones Centre, University of Cambridge, Trinity Ln, Cambridge, CB2 1TN, UK}

\author{Sandip De}%
\affiliation{%
BASF SE, Group Research, Carl-Bosch-Straße 38, 67056 Ludwigshafen, Germany
}%

\author{Gábor Csányi}
\affiliation{Max Planck Institute for Polymer Research, Ackermannweg 10, Mainz, Germany}
\affiliation{Engineering Laboratory, University of Cambridge, Trumpington St., Cambridge, UK}
\affiliation{Lennard-Jones Centre, University of Cambridge, Trinity Ln, Cambridge, CB2 1TN, UK}

\begin{abstract}
Machine learned interatomic potentials (MLIPs) have enabled atomistic simulations with \textit{ab initio} accuracy for a fraction of the computational cost.
However, many widely used MLIPs are short-ranged and do not accurately capture long-ranged electrostatic interactions.
At interfaces with polar solvents, such as water, this deficiency can drive unphysical long-distance dipolar alignment far away from the interface.
Here we reveal that neglecting long-ranged physics leads to spurious metallization of the water layer due to artificially large fluctuations of the total solvent dipole, 
similar to the electron rearrangement observed to prevent polar catastrophes at polar interfaces.
This metallization is eliminated in MLIPs that explicitly include long-ranged electrostatics.
Our results showcase a fundamental flaw of short-ranged MLIPs, highlighting that long-ranged electrostatics are essential for studying systems with a polar-liquid component, 
especially if one is interested in electronic properties.
\end{abstract}

\maketitle

    Machine learned interatomic potentials (MLIPs) are transforming how atomistic simulations are performed  \cite{behler_generalized_2007,bartok_gaussian_2010}.
    MLIPs allow for large scale, ab-initio quality molecular dynamics (MD) simulations by learning a mapping directly from the atomic coordinates of a molecule or crystal structure to the Born-Oppenheimer potential energy. 
    Applications include obtaining atomic-scale understanding of reactive systems \cite{schaaf_accurate_2023, vitartas2026active_metad}, construction of phase diagrams \cite{willman2022machine_prBl, kapil2022first_nanowater_phase},
    and high-throughput workflows for materials discovery \cite{sivak2025discovering_prl, chen2024accelerating}.

    The most widely used MLIPs are short-ranged, predicting total energies by summing atom-wise contributions that depend only on the local atomic environment within a cutoff radius $r_{\text{cut}}$ of each atom. 
    A shortcoming of the locality approach is a reduced ability to describe long-ranged interactions. A prominent example are electrostatic interactions, which decay slowly as $1/r$ and are therefore difficult to learn for short-ranged models \cite{anstine2023machine_lrlmlip_review}.
    Message passing MLIPs like MACE \cite{batatia_mace_2022,musaelian_learning_2023,bochkarev_graph_2024,fu2025learning_esen} propagate information from atoms beyond the radial cutoff resulting in an extended effective receptive field $r_{\text{eff}}$ for reduced computational cost, but remain fundamentally short-sighted.
    
    Consequently, considerable efforts have been devoted to introduce explicit electrostatics into MLIPs.
    One of the first approaches was to predict atomic charges from local environments, which are used to calculate a Coulomb energy~\cite{artrith_high-dimensional_2011,morawietz_neural_2012}.
    Subsequent developments include using localized Wannier function centres \cite{zhang2022deep_potential,gao_self-consistent_2022}, global charge equilibration schemes \cite{ghasemi2015interatomic_cent,ko2021fourth_4gnnp,staacke_kernel_2022} and long-ranged descriptors \cite{grisafi2019incorporating_lode1,huguenin2023physics_lode2}, amongst others \cite{unke2019physnet,xie_incorporating_2020,kabylda2025molecular_so3lr,cheng_latent_2025,bergmann_machine_2025}.

     Capturing electrostatic interactions is crucial for modelling polar solvents, like water, at interfaces.
     The environmental and industrial importance of these systems
     has driven an increasing number of computational studies, as their atomic-scale complexities are hard to resolve experimentally \cite{rossmeisl2008modeling_elec_sol_liquid_interface,
     carrasco2012molecular,
     gross2014water,
     björneholm2016water,
     quaranta_proton-transfer_2017,
     ngouana2017atomistic, levell2024emerging_hetelcat, 
     ghosh_contribution_2025,
     tian2025electrochemical_agh2o,
     advincula2025protons}. 
    While previous work has shown that short-ranged (SR) potentials can faithfully reproduce most properties of homogeneous bulk water \cite{morawietz2016van_rpbedens, yue_when_2021}, for interfacial systems that posses broken translational symmetry in one direction, they fail to correctly describe the depolarising forces arising from large spontaneous molecular dipole alignments.
    This leads to unphysical alignment of molecules across the length of the simulation cell, the extent of which depends upon the degree of interfacial ordering \cite{feller1996effect-truncation-dipole-jpc,rodgers2008interplay-hbond-dipole-nas,cox_dielectric_2020}.
    Whilst this alignment has been previously reported for SR-MLIPs
    \cite{niblett_learning_2021,gao_self-consistent_2022,cheng_latent_2025,grasselli_long-range_2026}, its implications for the solvent electronic structure remain unknown.
    In this letter, we find that short-ranged MLIPs produce solvent configurations which look reasonable, but do not make sense in electronic structure calculations. We also demonstrate this is an inevitable failure mode for all short-ranged MLIPs.
    


\begin{figure}
    \centering
    \includegraphics[width=1.0\linewidth]{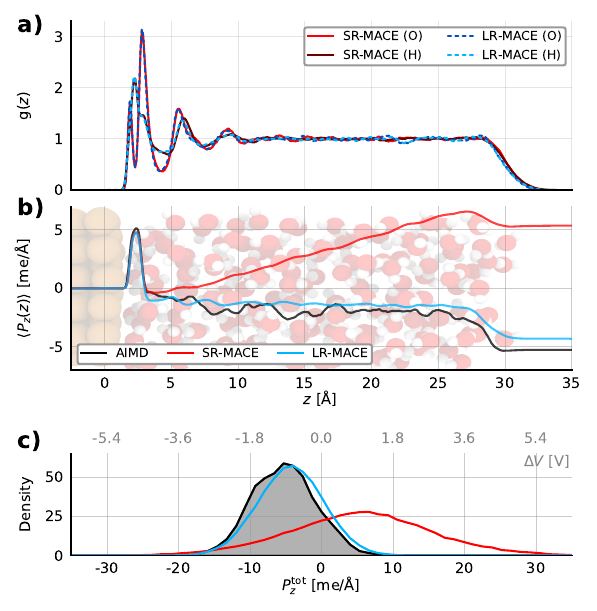}
    \caption{
    a) Density profile of water atoms above copper surface obtained from MD.  
    b) Integrated mean dipole per unit area of the water layer $\langle P_z(z) \rangle$. 
    Simulation setup shown in background.
    c) Distribution of total per unit area dipole $P^\text{tot}_z$. Equivalent potential difference $\Delta V$ shown above.}
    \label{fig:1:aimd_profiles}
\end{figure}

Throughout, we will compare the normal MACE architecture to a long-ranged (LR) MACE model with explicit electrostatics.
Atomic charges are obtained using a split-charge approach \cite{nistor2006generalization-split-charge}, starting from formal atomic oxidation states and then redistributing charge between neighbouring atoms based on the chemical environment, which ensures charge conservation.
Higher-order atomic multipoles are also predicted based on the local atomic environment. 
These charges and multipoles give a coulomb energy which is summed with the normal MACE model for the local energy to yield total energy predictions, with all parameters trained simultaneously.
Further details are contained in the supplementary material (SM) \cite{Supplementry_material_this}.


We begin by comparing the performance of MLIPs when simulating copper-water interfaces. 
Such metal-water interfaces are prevalent in electrochemical applications and heterogeneous catalysis \cite{seh_combining_2017,gonella_water_2021,omranpour_machine_2025}. 
We perform simulations of the interfacial system shown in Figure~\ref{fig:1:aimd_profiles}b using a SR-MACE and LR-MACE trained on \textit{ab initio} molecular dynamics (AIMD) \cite{marx-hutter2009ab-aimd} reference data, using density functional theory (DFT) \cite{hohenberg1964inhomogeneous, kohn1965self} with the PBE-D3 exchange-correlation functional \cite{perdew1996generalized_pbe,grimme2010consistent_d3}.
Both models achieve similar single point validation errors (Table~\ref{tab:S:aimd_models_cuw}), and produce near identical atom density profiles of the water, shown in Figure~\ref{fig:1:aimd_profiles}a.

By looking at the right quantity, however, one can find large differences between these models. Consider the cumulative dipole moment per unit area $\boldsymbol{P}(z)$ of the water layer, estimated by summing the dipole moment $\boldsymbol{m}_i$ of all molecules below $z$:

\begin{equation}
    \boldsymbol{P}(z)= \frac{1}{A} \sum_{\substack{i \in \text{O}\\ z_i<z}}^{N_\text{mols}}  \ 
    \boldsymbol{m}_i = \frac{1}{A} \sum_{\substack{i \in \text{O}\\ z_i<z}}^{N_\text{mols}}  \ 
    \sum_{\substack{j \in \text{H}_i\\}}  \frac{q}{2} (\boldsymbol{r}_{j} - \boldsymbol{r}_{i})
    \label{eq:dipole_calc_main}
\end{equation}

where $A$ is the surface area, $\text{H}_i$ denotes all Hydrogen atoms bonded to Oxygen $i$, and $q$ is a fixed partial charge of $0.5562\,e$, fitted to approximately reproduce the total DFT dipole of slabs of liquid water (Figure \ref{fig:S:water_slab_dipole_jump}).  
We emphasise this partial charge is unrelated to the LR-MACE architecture, and equation \eqref{eq:dipole_calc_main} is just a simple way to estimate dipolar alignment of any configuration, regardless of which model or method generated the trajectory.
Focusing on the $z$ component, the approximate total dipole moment of the entire water layer is $P^\text{tot}_z=P_z(z\rightarrow \infty)$.
Further details are provided in the SM \cite{Supplementry_material_this}.

\begin{figure*}
    \centering
    \includegraphics[width=1.0\linewidth]{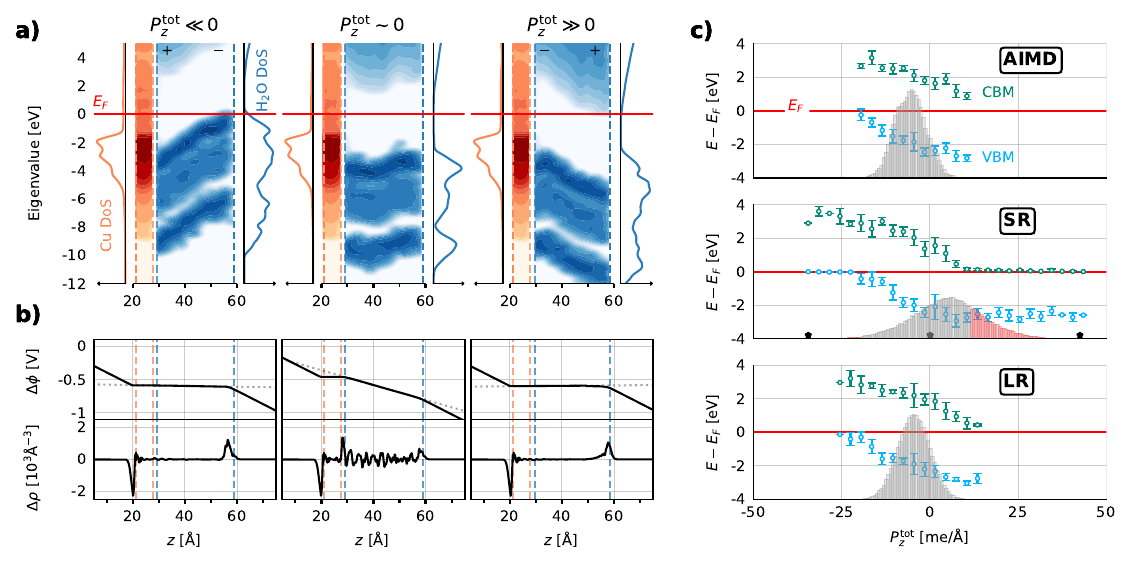}
    \caption{
    a) The electronic band structure under no applied field 
    for copper-water interfaces
    taken from an MD simulation ran using the SR-MACE described in the text. The shaded part in each plot is the PDoS versus $z$ for copper (orange) and water (blue). 
    Dashed lines represent the highest and lowest atom in each region. 
    The total DoSs are shown either side.
     b) Difference in electrostatic potential $\Delta \phi$ and electron density $\Delta \rho$ with $z$ between applied fields of 0 and $0.02\, \text{V/\AA}$ \space in the $z$ direction. The dotted grey line shows the gradient of $\Delta \phi$ in the water region.
     c) The VBM and CBM against binned dipole for AIMD, LR-MACE and SR-MACE simulations. The dipole distribution is shown in grey, with regions of breakdown shaded in red. Stars represent the configurations used in panels a/b.
    }
    \label{fig:1:pdos}
\end{figure*}

Figure~\ref{fig:1:aimd_profiles}b shows the mean of $P_z(z)$ across all samples from MD, and panel c shows the distribution of $P^\text{tot}_z$. 
In all cases, including AIMD, the average $\langle P^\text{tot}_z \rangle$ is non-zero, which can be attributed to interfacial ordering of water.
In AIMD, the cumulative dipole remains roughly constant in the centre of the water layer because the molecules are randomly oriented.
The SR-MACE exhibits a substantially larger variance $\sigma_{P^\text{tot}_z}^2$ and steadily increasing $\langle P_z(z) \rangle$, indicating unphysical molecular alignment across the entire slab.
The SM shows distributions from short-ranged foundation MLIPs \cite{batatia_foundation_2024}, which have even more pathological distributions \cite{Supplementry_material_this}.
In a slab geometry, the potential difference $\Delta V$ across the slab is proportional to the total dipole via $\Delta V=|P^\text{tot}_z|/\epsilon_0$ \cite{bengtsson_dipole_1999}.
The larger fluctuations of $P^\text{tot}_z$ in the SR-MLIP-MD therefore imply large voltage differences within the water layer, which---as we will show---greatly alters the electronic structure. 
In contrast, LR-MACE-MD closely matches the AIMD reference, with remaining differences arising from finite sampling statistics.





We now assess how artefacts in SR-MACE-MD impact the electronic structure for three sample configurations from the MD trajectory.
In Figure~\ref{fig:1:pdos}a, the three panels show the $z$-resolved projected density of states (PDoS) for configurations with large negative ($-34.5\,\text{m}e\text{/\AA}$), small ($\approx0.1\,\text{m}e\text{/\AA}$) and large positive ($42.5\,\text{m}e\text{/\AA}$) values of $P^\text{tot}_z$.
The copper and water density of states (DoS) are also shown on each side. 
In all cases, the Fermi level $E_F$ is fixed by the metal.

For $P^\text{tot}_z \approx 0$, $E_F$ lies in the gap between the water valence and conduction bands, as expected for an insulator \cite{creazzo_waterair_2024}. 
For large $|P^\text{tot}_z|$ however, the dipole creates an electric field across the water layer, causing the water bands to tilt.
For large negative $P^\text{tot}_z$,
the valence band maximum (VBM) intersects $E_F$ at the water-vacuum interface, and for large positive $P^\text{tot}_z$ the conduction band minimum (CBM) intersects $E_F$.
Similar erroneous band tilting has been observed in narrow ionic slabs, slabs with unrelaxed surface terminations \cite{meyer_density-functional_2003} and in a recent MLIP study of water under applied bias \cite{bergmann2026erasing_ref2}.


A necessary and striking consequence of this band bending is that electrons can flow between the water and metal. 
This can be seen by applying a small electric field of $0.02\,\text{V/\AA}$ in the $z$ direction during DFT, and looking at the change in electrostatic potential $\Delta \phi(r)$ and electron density $\Delta \rho$, as shown in Figure~\ref{fig:1:pdos}b.
For $P^\text{tot}_z \approx 0$, $\Delta \phi$ has a gradient in the vacuum region equal to the applied electric field strength and zero gradient in the conducting metal.
In the aqueous region $\Delta \phi$ has a smaller gradient, due to water's (electronic) dielectric screening. 
$\Delta \rho$ is zero in the metal centre, with peaks at either end from surface charge build-up \cite{price_molecular_1995}. In the water region $\Delta\rho$ exhibits structure at molecular length scales due to charge redistribution within molecules. 
For large $|P^\text{tot}_z|$, qualitatively different behaviour is observed: the water now displays metallic character, with flat $\Delta \phi$ throughout and charge redistribution no longer localised within molecules.
As electrons can move from the water-air interface into the metal, no charge builds up at this interface. 


We can analyse the extent to which this ``false metallization'' occurs by assessing the fraction of configurations sampled during MD where the water VBM or CBM coincide with the Fermi level.
Figure~\ref{fig:1:pdos}c quantifies this, showing how the VBM and CBM change with $P^\text{tot}_z$ for AIMD, LR-MACE, and SR-MACE simulations.
The distribution of $P_z^\text{tot}$ is plotted in the background, with regions of metallization marked in red. 
While AIMD exhibits negligible amount of metallization, SR-MACE's broadened dipole distribution leads to metallization in $\approx26\%$ of sampled structures.
Conversely, LR-MACE closely matches the AIMD distribution.
To summarise, while the validation errors and atom density profiles of the SR and LR models are near identical, many of the structures sampled from SR-MACE MD are fundamentally wrong, since the water is conducting in electronic structure calculations.

Several additional points merit discussion.
First, whilst water does undergo dielectric breakdown under large applied fields \cite{price_molecular_1995}, such breakdown should not occur spontaneously as observed for SR-MLIPs.
Second, PBE generally underestimates band gaps due to electron self-interaction \cite{chen2016ab_watergappbe}, potentially making water more susceptible to metallization. 
However, SR-MLIPs exhibit far more frequent metallization than AIMD.
Finally, AIMD studies suggest that excess electron charge on metals can ''spill over'' into water orbitals within $\approx5\,\text{\AA}$ \space from the interface \cite{li_electron_2025}. 
Unlike this spillover, which is physically reasonable, we see spurious metallization across the entire aqueous layer, which masks genuine near-interfacial effects and corrupts calculations of the extended electrical double layer and related properties such as capacitance.

\begin{figure*}
    \centering
    \includegraphics[width=1.0\linewidth]{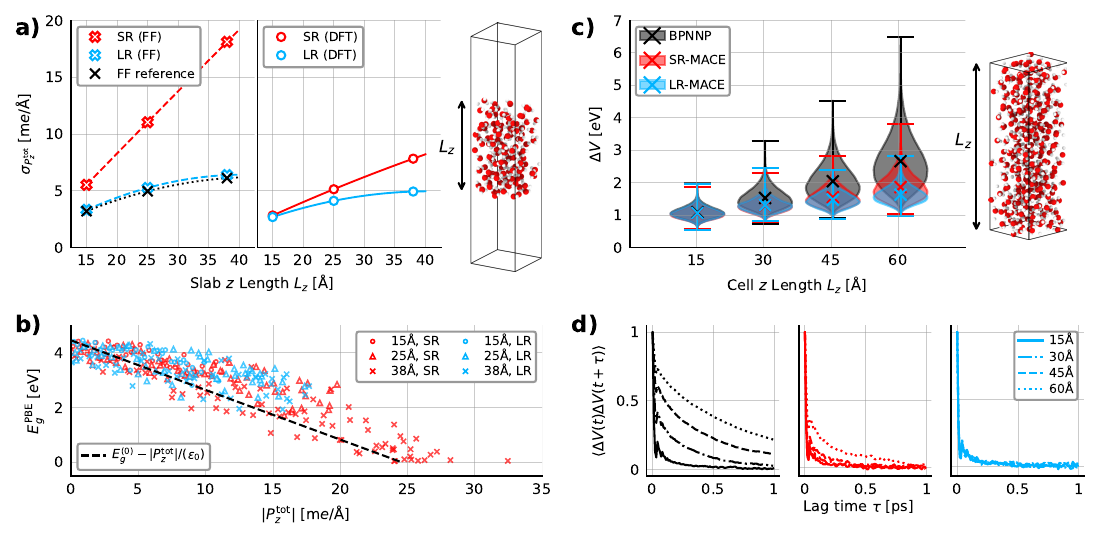}
    \caption{
            a) Standard deviation of the total dipole $\sigma_{P^\text{tot}_z}$ versus water-slab thickness $L_z$ for different potentials.
            b) PBE band gap $E_g^\text{PBE}$ versus $|P^\text{tot}_z|$. The black dashed line shows the theoretical change in potential difference where $E_g^{(0)}=4.45$.
            c) Violin plots of the cell potential difference $\Delta V$  of bulk water cells with $z$ length $L_z$ from MD using different MLIPs. 
            d) Autocorrelation functions of the change in $\Delta V$.}
    \label{fig:2:water_fig}
\end{figure*}

At this point one may question if other aqueous systems become metallized.
To answer this we benchmark model performance on the simplest aqueous interface: slabs of liquid water in vacuum. 
We trained SR- and LR-MACE models on both a revPBE0-D3 dataset \cite{schran_committee_2020} and to a classical force-field (FF) labelled dataset, and ran MD simulations of water slabs of varying widths.
The standard deviation of the total dipole $\sigma_{P^\text{tot}_z}$ was computed for each simulation, as shown in Figure~\ref{fig:2:water_fig}a.
For slabs with width $\approx15\,\text{\AA}$, the SR and LR models produce comparable fluctuations. As slab thickness increases, $\sigma_{P^\text{tot}_z}$ increases significantly for the SR-MLIP. 
Since SR-MACE has no strong long-ranged interactions, dipole moments decorrelate beyond the effective cutoff and the total dipole behaves like a random walk. Thus, its variance grows linearly with the number of molecules.
In Figure \ref{fig:2:water_fig}b, we see that configurations sampled from the thickest slab with large $|P^\text{tot}_z|$ have zero band gap using PBE ($E^{\text{PBE}}_g$).
Electronic structure calculations (Figure~\ref{fig:S:water_slab_pdos}) confirm the occurrence of metallization in these structures \cite{Supplementry_material_this}.
Similar metallization in response to increasing thickness is observed at polar/non-polar ionic interfaces to avoid ``polar catastrophes'' from a diverging total dipole moment  \cite{noguera_polar_2000, nakagawa2006some-interfaces-cannot-sharp, hu_kaolinite_2010}. 
Though the relief mechanism is similar, the metallization here arises as a simulation artefact due to model deficiencies, as opposed to being a genuine physical effect.
In comparison, $\sigma_{P^\text{tot}_z}$ increases slowly and plateaus for the LR-MLIP, closely matching the FF reference and producing no configurations in which $E_g=0$.

We further extend our analysis to see if false metallization occurs even in fully periodic systems. 
We simulate boxes of water, elongating one dimension $L_z$. 
Alongside our water MACE models, MD was performed using a Behler-Parrinello neural network potential (BPNNP) from reference \cite{schran_committee_2020}. 
Unlike the non-periodic case, there is no well defined dipole moment that will correlate with metallization \cite{king-smith_theory_1993}.
One instead can estimate whether metallization will occur by computing the approximate electrostatic potential in the cell and examining the difference between the highest and lowest values, denoted $\Delta V$. 
The electrostatic potential causes a bending of the bands, similar to Figure \ref{fig:1:pdos}. 
If $\Delta V$ exceeds the band gap, metallization will occur.
$\Delta V$ in each frame was estimated by using fixed partial charges, as in equation \eqref{eq:dipole_calc_main}, and computing the potential throughout the cell by assigning a spherical Gaussian charge distribution to each atom.

The distributions of $\Delta V$, shown in Figure \ref{fig:2:water_fig}c, reveal a general trend that the mean of $\Delta V$ increases with $L_z$.
For $L_z=15-30\,\text{\AA}$, all models produce similar distributions.
When $L_z \gg r_{\text{eff}}$, the SR model's mean and variance increase beyond that of the LR model. Deviation occurs when $L_z \gtrsim 4r_{\text{eff}}$, likely due to the correlation of the atomic forces in SR-MLIPs over this distance.
The BPNNP therefore diverges for smaller $L_z$ due to its roughly two times smaller receptive field compared to the SR-MACE.
The PDoS for the highest $\Delta V$ configuration displays closing or significant narrowing of the band gap even for a hybrid functional (Figure~\ref{fig:S:rod_water_pdos}).
We therefore expect that larger systems simulated with SR-MLIPs will become metallic regardless of the reference method used \cite{Supplementry_material_this}.

    
Analysis of the autocorrelation functions (ACFs) of $\Delta V$ in Figure~\ref{fig:2:water_fig}d reveals similar trends.
For the BPNNP, the ACF decays significantly slower as $L_z$ increases.
Comparatively, the SR-MACE decorrelation times are similar for small $L_z$, but differ when $L_z=60\,\text{\AA}$. 
The LR-MACE ACFs agree irrespective of $L_z$. 
We also analysed the ACFs of $P^\text{tot}_z$ for the copper-water interfaces (Figure~\ref{fig:2:acf_size}a).
We find the LR-MACE closely resembles the AIMD ACF whilst the SR-MACE decays an order of magnitude slower. 
Taken together, these results show that SR-MLIPs also incorrectly reproduce dynamical properties, whereas LR-MACE remains robust.


Given that our results indicate SR-MLIPs give similar performance to LR-MLIPS for small systems sizes, one may wonder if SR-MLIPs can avoid metallization during practical use.
In the SM (Figure~\ref{fig:2:acf_size}b) we have explored two commonplace approaches for improving the performance of SR-MACE for the copper-water interfacial system \cite{Supplementry_material_this}. 
We tried (i) increasing the receptive field of the model by using more message passing layers, and (ii) enlarging the training set through iterative learning, to include structures with $P_z^\text{tot}$ not encountered in AIMD. 
Our experiments show that by using very large receptive fields, combined with iterative learning, short-ranged models can approach the properties of AIMD. 
It was found, however, that even in this limit a small LR-MACE trained only on AIMD performed best. 
Furthermore, as the cumulative dipole grows gradually across the water layer, false metallization is expected to become more frequent as system size increases, and any apparent model avoidance may not transfer to larger systems, as seen in Figure \ref{fig:2:water_fig}.
Indeed, upon extending the water layer of the copper-water interface, we found that even applying all the above mentioned tricks is insufficient to train an adequate SR-MLIP (Figure \ref{fig:S:A90}).

We end our analysis by considering other polar systems.
The water slab represents the case of two 'disordered' interfaces, the copper-water-vacuum system one 'ordered', one 'disordered' interface.
Metallization similarly occurs in the case of a fully periodic copper/water system (with no vacuum region), featuring two 'ordered' interfaces (Figure~\ref{fig:S:closed_pdos}) \cite{Supplementry_material_this}.
$P^\text{tot}_z$ broadening was further observed in SR-MLIPs simulations of the \ce{TiO2 rutile (110)} water interface (Figure~\ref{fig:S:tio2}), generalising these results beyond metallic and vacuum interfaces \cite{Supplementry_material_this}.
Finally, cubic bulk water cells follow similar trends in potential difference distributions to the aforementioned elongated cells (Figure~\ref{fig:S:cubic_water_V}) \cite{Supplementry_material_this}.

In conclusion, we have shown that deficiencies in describing long-ranged electrostatic information in SR-MLIPs leads to unphysical large spatial variations in potential difference, which causes false metallization of systems with a polar liquid component, be they interfacial or bulk.
LR-MLIPs alleviate this issue by explicitly modelling electrostatics. 
Furthermore, SR-MLIPs produce sluggish depolarisation dynamics, with longer correlation times compared to LR-MLIPs and AIMD references.
SR-MLIPs with a large enough receptive field can mitigate some artifacts, but inevitably will fail for large enough systems.
Considering the wider impact of our findings, we suggest SR-MLIPs may produce incorrect observables which require correct dipolar behaviour, for example transport coefficients, work functions and capacitances.
Studies of heterogeneous catalysis where the electric double layer affects how molecules react at surfaces may also suffer.

Finally, the current generation of materials-focussed foundation MLIPs are all short-ranged. 
Our results show that this limits their accuracy for a broad class of systems.
Looking forward, we expect that incorporating explicit electrostatics is the next step towards a truly universal MLIP for atomistic modelling.

\subsection{Author Contribution Statement}
I.J.P: Data curation, Formal analysis, Investigation, Visualization, Writing – original draft, Writing – review \& editing. M.J.H: Data curation, Formal analysis, Investigation, Visualization, Writing – review \& editing. W.J.B: Conceptualization, Formal analysis, Investigation, Methodology, Software, Supervision, Writing – review \& editing. S.H: Investigation, Writing – review \& editing. S.G: Supervision, Writing – review \& editing. K.D.F: Conceptualization, Supervision, Writing – review \& editing. A.M: Conceptualization, Funding acquisition, Project administration, Supervision, Writing – review \& editing. C.S: Conceptualization, Methodology, Project administration, Resources, Supervision, Writing – review \& editing. S.D: Conceptualization, Funding acquisition, Methodology, Project administration, Supervision, Writing – review \& editing. G.C: Conceptualization, Funding acquisition, Methodology, Project administration, Resources, Supervision, Writing – review \& editing.

\subsection{Acknowledgements}

\begin{acknowledgments}
I.J.P. acknowledges BASF SE for corporate funding.  W.J.B. thanks the Cambridge university engineering department for funding through the Ashby postdoctoral fellowship. M.J.H. acknowledges the Cambridge Trust for funding. A.M. thanks the European Union for support under the ``n-AQUA'' European Research Council project (Grant No. 101071937). The author's would like to acknowledge useful discussions with Ansgar Schaefer and Tiago Joao Ferreira Goncalves. C.S. acknowledges insightful discussion about the limit of short-ranged models with Steve Cox. 

We acknowledge the use of ARCHER2 UK National Supercomputing Service \cite{beckett2024archer2}. This work also was performed using resources provided by the Cambridge Service for Data Driven Discovery (CSD3) operated by the University of Cambridge Research Computing Service, provided by Dell EMC and Intel using Tier-2 funding from the Engineering and Physical Sciences Research Council (capital grant EP/T022159/1), and DiRAC funding from the Science and Technology Facilities Council. We acknowledge the use of resources provided by the Isambard 3 Tier-2 HPC Facility. Isambard 3 is hosted by the University of Bristol and operated by the GW4 Alliance (https://gw4.ac.uk) and is funded by UK Research and Innovation; and the Engineering and Physical Sciences Research Council [EP/X039137/1]. We acknowledge the EuroHPC Joint Undertaking for awarding the project ID EHPC-REG-2024R02-130 access to the EuroHPC supercomputer LEONARDO, hosted by CINECA (Italy) and the LEONARDO consortium. We acknowledge the use of AI tools (Claude) to polish aspects of the text.
\end{acknowledgments}

\subsection{Conflict of interests}
G.C. is a partner in Symmetric Group LLP that licenses force fields commercially and also has equity interest in Ångström AI.

\subsection{Data Availability}
Data is currently available upon request. Models, training sets, input files and analysis scripts will be made public at a later date.

\subsection{Supplementary Material}
In the supplementary material we provide computational details, model architecture details, derivations, further supporting simulations, analysis of other aqueous systems, and a more detailed breakdown of short-ranged model optimisation procedures.

\bibliographystyle{apsrev4-2}
\bibliography{ref2,ijp30_references}


\newpage
\clearpage

\setcounter{section}{0}
\setcounter{equation}{0}
\setcounter{figure}{0}
\setcounter{table}{0}
\setcounter{page}{1}

\renewcommand{\thesection}{S\arabic{section}}
\renewcommand{\theequation}{S\arabic{equation}}
\renewcommand{\thefigure}{S\arabic{figure}}
\renewcommand{\thetable}{S\arabic{table}}
\renewcommand{\thepage}{S\arabic{page}}

\title{Supporting Material for: \mytitle}
{\maketitle}



\section{Computational Methods}

We find it useful first to define some terminology that will be used throughout the supplementary material, shown in Table~\ref{tab:notation}. Unless stated otherwise in the text, symbols correspond to the definition below.
\begin{table}[h!]
    \centering
    \caption{Frequently used symbols used and their definition}
    \begin{tabular}{|c|l|}
        \hline
        Symbol & Definition \\
        \hline
        $K_{\text{channels}}$ & Number of channels in a MACE model \\
        $L$ & Maximum equivariance degree \\
        $N_\text{layers}$& Number of message passing layers\\
        $r_{\text{cut}}$ & Radial cutoff\\
        $r_{\text{eff}}$ & Effective radial cutoff\\
        $P_z(z)$ & $z$ component of cumulative dipole to height $z$\\
        $P_z^{\text{tot}}$ &  $z$ component of total dipole \\
        $\mathcal{NN}(i)$ & Nearest neighbours of $i$\\
        $\gamma $& Damping of Langevin thermostat.\\
        $E_g $& Band gap\\
        $\sigma_X$& Standard deviation of observable $X$\\
        \hline
    \end{tabular}
    
    \label{tab:notation}
\end{table}

\subsubsection{AIMD simulations}

AIMD was performed in CP2K \cite{kuhne2020CP2K} using the PBE functional \cite{perdew1996generalized_pbe} with the D3 dispersion correction \cite{grimme2010consistent_d3} applied.
For the plane wave basis set a cutoff of $500\, \text{Ry}$ was used.
The DZVP-MOLOPT-SR-GTH basis set was used with GTH-PBE pseudo potentials, only treating 11 electrons per Cu atom explicitly. 
We used gamma-point Brillouin zone sampling. 
The simulation setup consists of a 4-layer $5\times6$ Cu fcc(111) slab with a cell size of 
($12.617\times13.112\times58.403\, \text{\AA}$). 
The lattice parameter was optimised using the above DFT settings and found to be $3.56875\, \text{\AA}$.
165 water molecules were placed above the slab, corresponding to a $\approx30\, \text{\AA}$ \space thick layer.
AIMD was ran for $36\, \text{ps}$ at $330\, \text{K}$ to compensate for the well-known overstructured nature of PBE water, using a Nosé-Hoover thermostat \cite{nose1984unified,nose1984molecular} with chain length 4, a chain time constant of  $40\, \text{fs}$ and 2 multiple timesteps for the chain. 
A $1.0\, \text{fs}$ timestep in combination with deuterium masses for hydrogen atoms was used and the bottom 2 layers of the copper slab were kept fixed.

\subsubsection{Local-split-charge MACE model}

The local split-charge MACE architecture decomposes the total energy $E$ into a global long-range term $E^{\text{LR}}$ and a short-range term $E^{\text{SR}}$, which is a sum of atomic contributions. 
\begin{equation}
    E=E^{\text{LR}}+E^{\text{SR}}
\end{equation}
The atomic short-range energy is predicted using readouts on the local atomic features which are created during message passing \cite{batatia_mace_2022}, as in a normal MACE model.

To compute a long-range energy contribution, we adopt a transparent and physically motivated design. The model predicts a set of electric multipole moments, denoted $p_{i,lm}$, on each atom $i$, where $(lm)$ is an angular momentum index tuple. $l=0$ corresponds to atomic charges, $l=1$ to atomic dipole moments, and one can find a linear mapping from higher $l$'s to higher order Cartesian multipole moments.
A smooth charge density $\rho(\boldsymbol{r})$ can then be defined by combining the multipole moments with Gaussian type orbital basis functions $\phi_{lm}(\boldsymbol{r})$:
\begin{equation}
    \rho(\boldsymbol{r})=  \sum_{i,lm} p_{i,lm}\phi_{lm}(\boldsymbol{r}-\boldsymbol{r}_i)
\end{equation}
The explicit form of $\phi_{lm}$ is:
\begin{equation}
    \phi_{lm}(\boldsymbol{x})= C Y_{lm}(\hat{\boldsymbol{x}}) x^l e^{-x^2/2\sigma^2}
\end{equation}
in which $C$ is a normalization constant such that $\phi_{lm}(\boldsymbol{r})$ has its $lm$ multipole moment equal to 1.
For instance, $\phi_{00}$ has a charge of 1, $\phi_{1m}$ has a dipole moment of 1, and so on.
The long-range energy contribution is then simply defined as:
\begin{align*}
    E_{LR} = \frac{1}{2}\iint \frac{\rho(\boldsymbol{r}')\rho(\boldsymbol{r})}{4\pi \epsilon_0 |\boldsymbol{r}-\boldsymbol{r}'|} d \boldsymbol{r} d\boldsymbol{r'}
\end{align*}
The use of Gaussian type orbitals allows for simple, explicit evaluation of the total electrostatic energy in both periodic and open boundary conditions \cite{batatia2026mace-polar}. In this paper, all our simulations are performed either in fully periodic boundary conditions or in a slab geometry. For the fully periodic case we evaluate the above integral in a way similar to Ewald summation with tin foil boundary conditions, while in the slab case we apply the dipole correction as is done in DFT calculations of slab geometries. Details of the electrostatics implementation can be found in reference \cite{batatia2026mace-polar}. In this paper, all models only use atomic charges and atomic dipoles ($l$=0,1).

The architecture predicts $l=1$ multipoles directly as a linear function of the local node features.
Specifically, if the MACE node features after layer $s$ are denoted $h_{i,klm}^{(s)}$ (as in reference \cite{kovacs_evaluation_2023}), the atomic dipoles are predicted as:
\begin{align*}
    p_{i,1m} = \sum_{s=1}^S \sum_{k}^{K_\text{channels}} w_{k}^{(s)} h_{i,k1m}^{(s)}
\end{align*}
In which $S$ is the total number of layers. A slightly different procedure is used to predict atomic charges in order to ensure conservation of total charge.
Instead of directly predicting the total charge on an atom, the model instead predicts a charge transfer $p_{ij}$ between each pair of neighbouring atoms $i$ and $j$, where two atoms are defined as neighbouring when atom $j$ is within the local cutoff of atom $i$.
The charge on atom $i$ is then computed as:
\begin{equation}
    q_i = q^{(0)}_i + \sum_{j \in \mathcal{NN}(i)} (p_{ij}-p_{ji})
\end{equation}
 Where $q^{(0)}_i$ is a predefined formal oxidation state that is fixed throughout the simulation. In this case, the values $p_{ij}$ are computed as functions of the node features of both atoms, and their relative displacement. The values of $q^{(0)}$ used in this study are shown in Table~\ref{tab:S:ox_states}.
This scheme is referred to as `local split charge', to emphasize the connection to split charge polarization models \cite{Jensen2023UnifyingModels, nistor2006generalization-split-charge}.
This framework also allows one to define the total polarisation vector of the system $\boldsymbol{P}^{\text{MLIP}}$ as:
\begin{equation}
\begin{split}
    \boldsymbol{P}^{\text{MLIP}} &=\sum_{i=1}^N (\boldsymbol{p}_i + q^{(0)}_i\boldsymbol{r}_i)  + \\
    &\qquad  \sum_{i=1}^N\sum_{j \in \mathcal{NN}(i)}(p_{ij}\boldsymbol{r}_{ij}+p_{ji}\boldsymbol{r}_{ji})
\end{split}
\end{equation}
This polarization vector will be invariant to translation and will change by the correct amount if an atom is pulled through the unit cell boundary to replace one of its periodic images (provided $q^{(0)}_i$ corresponds to a formal oxidation state). 
The local split charge model can be trained on reference electronic structure values such as Hirshfeld charges \cite{hirshfeld1977bonded} or Hartree multipoles, though sensible model behaviour could also be obtained without training on any reference charges. Throughout this paper, the model is trained only on atomic forces and total energies, in the same way as the local MACE model. 
This is similar to the training method used in the latent Ewald sum method \cite{cheng_latent_2025}.

The prediction of higher order multipoles, use of formal oxidation states, and ability to predict a translation invariant $\boldsymbol{P}^{\text{MLIP}}$ for periodic systems differentiates our approach from others which directly predict charges.

\begin{table}[h]
    \centering
    \caption{Formal oxidation states $q^{(0)}$ used in the local split charge MACE, often referred to as LR-MACE, for each element included in this study.}
    \begin{tabular}{|cc|}
    \hline
        Element & $q^{(0)}$\\
    \hline
       Cu  & $0$\\
       H   & $+1$\\
       O   & $-2$\\
       Ti  & $+4$\\
    \hline
    \end{tabular}
    
    \label{tab:S:ox_states}
\end{table}

\subsubsection{Connection between total dipole and potential difference drop across a slab}

As discussed in reference \cite{bengtsson_dipole_1999}, if one considers a charge density $\rho_\text{av}(z)$ which is averaged across the $x$ and $y$ directions, the total drop in potential across such a slab is $\Delta V = 4\pi m$ in atomic units, or $\Delta V = m/\epsilon_0$ in SI units, where $m$ is the dipole moment of the slab per unit area. 

For charge densities which vary in $x$ and $y$, it is also pointed out in \cite{bengtsson_dipole_1999} that the impact on the electrostatic potential decays exponentially outside of the slab, and hence the formula can be applied to general charge densities. 
Thus, this approach was used in the present work to calculate the potential difference drop across our slabs.

\subsubsection{Estimation of dipole moments}

\begin{figure*}
    \centering
    \includegraphics[width=1\linewidth]{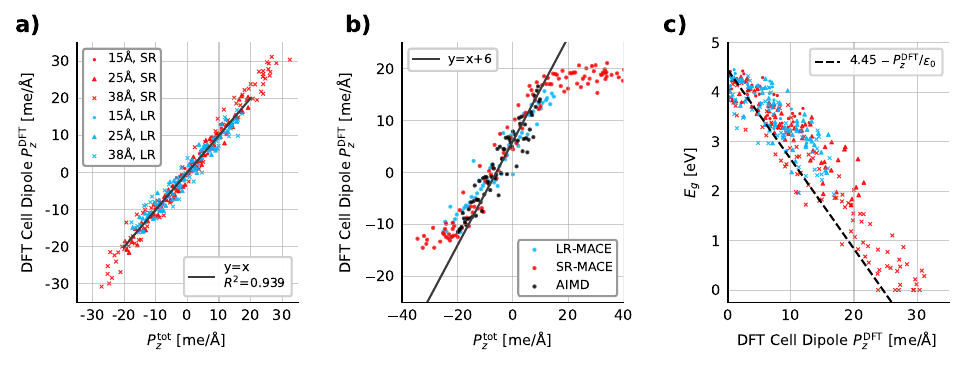}
    \caption{
    a) $P^{\text{tot}}_z$ versus DFT total cell dipole $P_z^{\text{DFT}}$ for configurations sampled from MD trajectories of water slabs of different thicknesses sampled using different MLIP architectures (see Figure~\ref{fig:2:water_fig}. The black line shows the correlation between the 2 metrics in the region used to fit $P_z^{\text{tot}}$.
    b) $P^{\text{tot}}_z$ of the water layer versus DFT total cell dipole for configurations sampled from MD simulations of copper-water interfaces. Sampled using different MLIP architectures and from AIMD. The black line shows the correlation between the 2 metrics, with a shift to account for the dipoles on the copper slab surfaces.
     c) $E_g$ versus DFT total cell dipole. Dashed black line shows the theoretical change in band gap with $P_z^{\text{DFT}}$. 
     Origin of $4.45\, e\text{V}$ was chosen based upon highest band gap sampled from DFT.
     Points use the same legend as panel a.
    }
    \label{fig:S:water_slab_dipole_jump}
\end{figure*}

Throughout the paper we employ an estimate of the total dipole based on fixed charges, as in equation \ref{eq:dipole_calc_main}, reproduced here:
\begin{align*}
    P_z^{\text{tot}} 
    &= \frac{1}{A} \sum_{\substack{i \in \text{O}\\ z_i<z}}^{N_\text{mols}}  \ 
    m_{i,z} = \frac{1}{A} \sum_{\substack{i \in \text{O}\\ z_i<z}}^{N_\text{mols}}  \ 
    \sum_{\substack{j \in \text{H}_i,O_i\\}}  q_j z_{j} \\
    &= \frac{1}{A} \sum_{\substack{i \in \text{O}\\ z_i<z}}^{N_\text{mols}}  \ 
    \sum_{\substack{j \in \text{H}_i\\}}  \frac{q}{2} (\boldsymbol{r}_{j} - \boldsymbol{r}_{i})
    \label{eq:dipole_calc_si}
\end{align*}
This definition necessitates well-defined molecular entities. Here $N_{\text{mols}}$ is equal to the number of oxygen atoms.
We assign a hydrogen atom to be in the same molecule as an oxygen atom if it is within $1.2\, \text{\AA}$ of the oxygen atom.
For systems where substantial dissociation occurs, the above formula breaks down, since fixed charges will not reproduce the correct formal charge of the resulting ionic constituents.
For the systems studied here, our AIMD results suggests that permanent dissociation should not occur (see Figure \ref{fig:S:dissociation}), making the approximation valid for physically reasonable systems (according to the reference method). 
The value of the partial oxygen charge $q$ was obtained by finding a best fit of the approximate dipole expression to the dipole obtained from DFT calculations, $P^{\text{DFT}}_z$, of the water slab system discussed in the main text.
This is shown in Figure~\ref{fig:S:water_slab_dipole_jump}a, which compares the actual DFT dipole with the approximation above for optimised charge values to provide the best correlation between the two.
The best value of $q$ was found to be $q=0.5562e$. When optimising $q$, we only used dipole values where $|P^{\text{DFT}}_z| < 20\, \text{m}e\text{V}$.
For a cell of $z$ dimension $L_z$, $P_z^{\text{DFT}}$ was calculated from the dipole correction gradient $\beta$ outputted from the DFT calculations in FHI-AIMS using the following formula
\begin{equation}
    P_z^{\text{DFT}}=\epsilon_0 \beta L_z 
\end{equation}
Comparing our estimate of the total cell dipole to $P_z^{\text{DFT}}$ for the copper-water interface, shown in Figure~\ref{fig:S:water_slab_dipole_jump}b, provides one further piece of evidence of metallization. 
One sees $P_z^{\text{tot}}$ correlates well with $P_z^{\text{DFT}}$ for smaller dipoles.
However, $P_z^{\text{DFT}}$ plateaus for large $P_z^{\text{tot}}$.
This is because electrons redistribute across the system to prevent the dipole increasing any further to avoid the high electrostatic energy cost, akin to the polar catastrophes mentioned in the main text.
Therefore our estimate of $P_z^\text{tot}$ holds only for non-metallised systems.
Explicitly, this includes those situations where electrons stay localised within water molecules.

\subsubsection{PDoS}

Calculations of the projected density of states (PDoS) and impact of applied field were performed using FHI-AIMS \cite{blum2009ab_fhiaims,yu2018_elsi_aims,havu2009efficient_aims_grid} using the PBE functional.
The intermediate basis set with a k-point grid of $4\times4\times1$ was used.
A Gaussian smearing of $0.025\, e\text{V}$ was chosen, and relativistic corrections were turned off.
A dipole correction was applied in the $z$ direction. 

The PDoS was calculated for 1001 energy values between -20 and $5\, e\text{V}$ (a spacing $\varepsilon$ of $0.025\, e\text{V}$) with a smearing of $0.01\, e\text{V}$. In each configuration, the eigenvalues are then shifted so $E_F=0$.

To estimate the energy eigenvalue of the VBM and CBM of the water layer --- $E_\text{VBM}$ and $E_\text{CBM}$ respectively --- as shown in Figure~\ref{fig:1:pdos}c, the following algorithm is employed:
\begin{itemize}
    \item Sum the PDoS for Hydrogen and Oxygen atoms that are greater than a chosen distance $z_{\text{max}}$ from the highest Copper atom in the slab. This is because the basis functions of the molecules immediately next to the copper slab overlap with those in the slab, which makes defining water-only states difficult.
    \item If $P^{\text{tot}}_z$ is negative, the bands should tilt upward (see the left hand panel in Figure \ref{fig:1:pdos}a), so find the first eigenvalue below $E_F+ W \varepsilon$ with total PDoS greater than a minimum threshold $X$, depending on the system's size.  
    This eigenvalue is $E_\text{VBM}$. 
    We start looking a multiple $W$ of the spacing $\varepsilon$ above the bands to account for bands that may cross the Fermi level if the system is metallized. Now find the first eigenvalue above $E_{\text{VBM}}+S\varepsilon$. This is $E_\text{CBM}$.
    $S\varepsilon$ ensures that different bands/orbitals are detected, as the bands are smeared slightly by a Gaussian during post processing. Thus is $S=0$ we would detect the same band.
    \item If $P^{\text{tot}}_z$ is positive, the bands should tilt downward (see the right hand panel in Figure \ref{fig:1:pdos}a), so find the first eigenvalue above $E_F- W \varepsilon$ with PDoS greater than $X$. 
    This eigenvalue is $E_\text{CBM}$. 
    Now find the first eigenvalue below $E_{\text{CBM}}-S\varepsilon$. This is $E_\text{VBM}$. $W$ and $S$ serve the same purpose as before.
    \item The band gap is calculated as $E_g=E_\text{CBM}-E_{\text{VBM}}$
\end{itemize} 
Here we used $z_{\text{max}}=2.5\, \text{\AA}, X=12, W=15$ and $S=50$.
In Figure~\ref{fig:1:pdos}c, we bin our predicted band gaps within a region of $-0.054$ to $0.054\, e\text{V}$ using a bin width of $0.03\, e\text{V}$, and plot the mean and standard deviation within each bin.

To calculate the effect of an applied field, DFT calculations on selected configurations were performed using a homogeneous field in the $z$ direction $E_z$ of 0.00 and $0.02\, \text{V/\AA}$. 
The potential of the system $\phi=v_\text{ext} + v_H$, where $v_\text{ext}$ and $v_H$ are the external and Hartree potentials respectively.
The difference in the potential $\Delta \phi$ is calculated as:

\begin{equation}
\begin{split}
\Delta \phi(z)
&=
\langle v_H(\boldsymbol{r},E_z=0.02) \\
&\qquad- v_H(\boldsymbol{r},E_z=0.0) - 0.02 z \rangle_{xy},
\end{split}
\end{equation}

where $\langle\cdots \rangle_{xy}$ represents an average over at a given $z$ value over $x$ and $y$ coordinates. 
The difference in the electron delta density $\rho$ is calculated as:

\begin{equation}
    \rho(z)=\langle \rho(\boldsymbol{r},E_z=0.02)-\rho(\boldsymbol{r},E_z=0.0)\rangle_{xy}
\end{equation}


\subsection{Copper-Water System}

\subsubsection{Model and Simulation Details }

MACE models were initially trained on a randomly sampled subset of the AIMD data.
These are the models shown in Figure \ref{fig:1:aimd_profiles}.
The SR-MACE model was trained from scratch using $r_\text{cut}=6.0\, \text{\AA}$, $N_{\text{layers}}=2$, $k=128$ and $L=1$.
These settings are the ones throughout the text except where stated otherwise. 
When training any model in the text, we take the checkpoint with the lowest loss for production runs.
500 configurations with a range of net dipoles were sampled, 125 of which were used for validation and the rest for training.
The single point test errors of the long- and short-ranged models are given in Table~\ref{tab:S:aimd_models_cuw}.

MD was ran for a total of $1.005\, \text{ns}$ via the Atomic Simulation Environment \cite{bahn2002ase,larsen2017ase} using a Langevin thermostat \cite{brunger1984stochastic_langevin} with a friction constant of $0.05\, \text{ps}^{-1}$.
The same temperature, timestep and masses as in the AIMD simulation were used. Frames were saved every $10\, \text{fs}$.
In all cases the first $5\, \text{ps}$ were discarded before analysis to allow for equilibration.

\begin{table}
    \centering
    \caption{Errors of models on forces and energies used for copper-water interface, trained on AIMD data.}
    \begin{tabular}{|c|cc|}
    \hline
        \shortstack{Model \\ name}  & \shortstack{RMSE($E$) \\ {[m$e$V/atom]}} & \shortstack{RMSE($F$) \\ {[m$e$V/\AA]}} \\
    \hline
        LR-MACE & 0.14 & 46.23 \\
        SR-MACE & 0.14 & 48.58 \\
        \hline
    \end{tabular}
    
    \label{tab:S:aimd_models_cuw}
\end{table}

\subsubsection{Performance of Foundation Models}

\begin{figure*}
    \centering
    \includegraphics[width=1.0\linewidth]{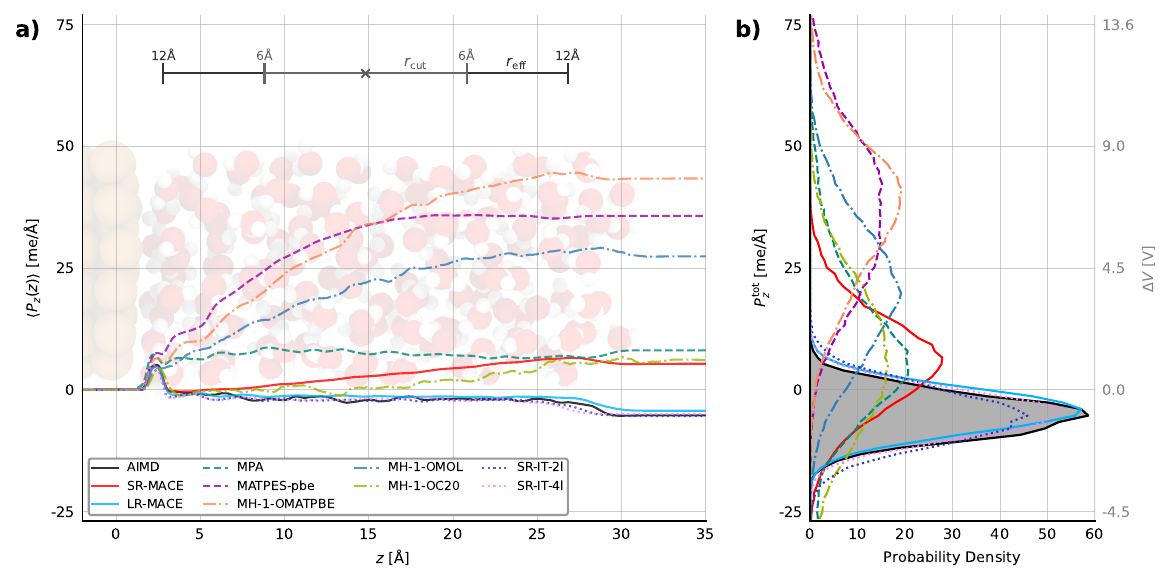}
    \caption{
    a) The integrated mean dipole per unit area of the water layer $\langle P_z(z) \rangle$ obtained from MD simulations using AIMD and different MLIP models. 
    As well as data in Figure~\ref{fig:1:aimd_profiles}, the MACE-MPA, MACE-MATPES-pbe and MACE-MH-1 foundation models are also included.
    Also show are are SR-MACE-$N$l, which refer to short-ranged MACE models trained on an iteratively trained dataset with $N$ message passing layers. Local model radial $r_\text{cut}$ and effective $r_\text{eff}$ cutoff for 2 layer models are indicated relative to a point marked by 'X'. 
    Simulation setup is shown in the background.
    b) Distribution of total per unit area dipole $P^\text{tot}_z$ obtained from MD. Equivalent potential difference $\Delta V$ shown to right. AIMD distribution is shaded to make clear it is the reference.
    }
    \label{fig:S:aimd_profiles_fm}
\end{figure*}

\begin{figure*}
    \centering
    \includegraphics[width=1\linewidth]{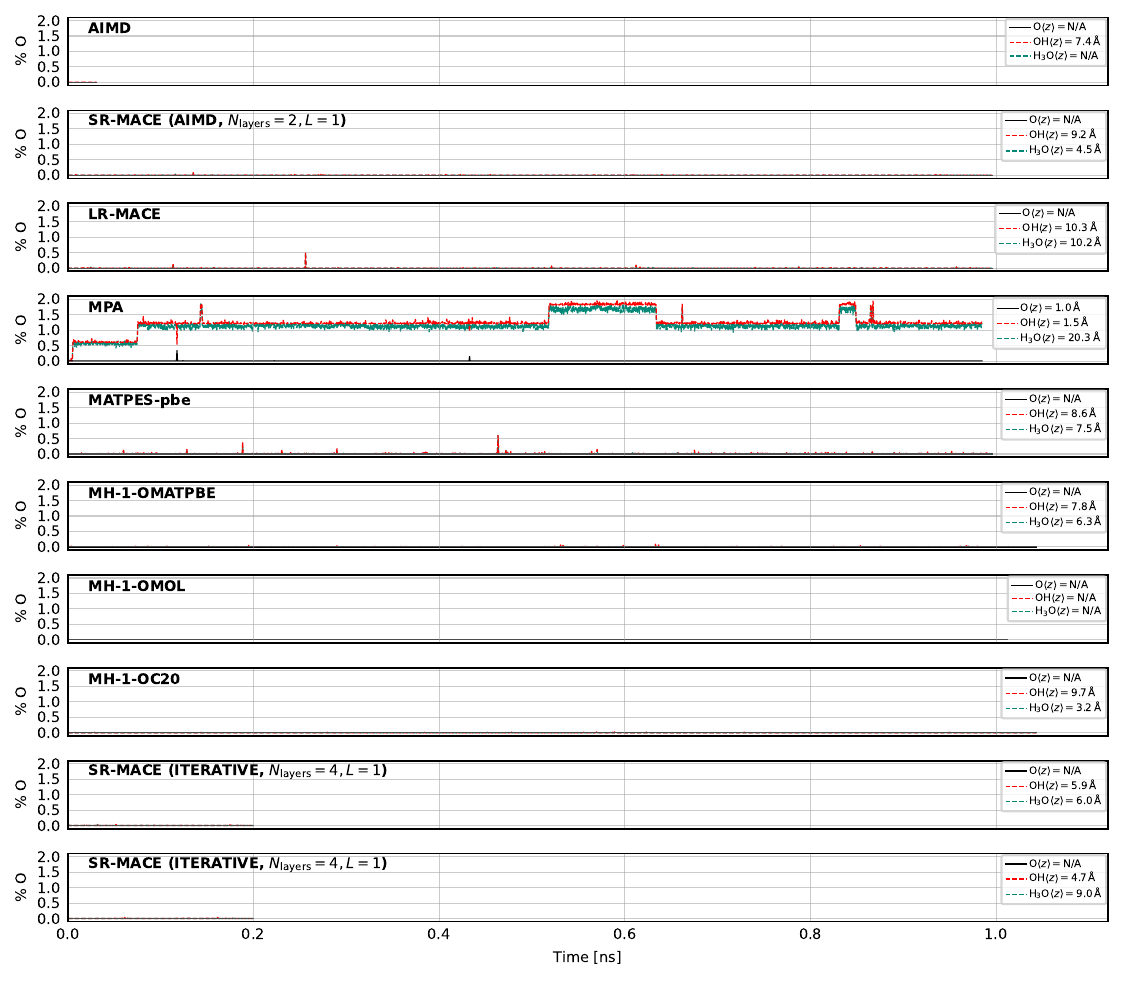}
    \caption{Percentage of total oxygens as dissociated species O, OH, and \ce{H3O} during MD simulation of copper-water interface using different potentials.
    Plotted as rolling average over $500\, \text{fs}$ window.
    Contains reference AIMD trajectory and all MLIPs shown in Figure~\ref{fig:S:aimd_profiles_fm}.
    Also shown is average position of species relative to surface. 
    Average position is given as N/A if species not observed.
    First $5\, \text{ps}$ were discarded in all cases.
    }
    \label{fig:S:dissociation}
\end{figure*}

Figure~\ref{fig:S:aimd_profiles_fm} reproduces the $P_z(z)$ distribution for the AIMD, SR-MACE and LR-MACE as shown in the main text, but further includes three MACE foundation models: MPA \cite{ghahremanpour2018alexandria}, MATPES-pbe \cite{kaplan2025foundational,batatia_foundation_2024} and MH-1 \cite{batatia_cross_2025}. 
Most of these foundation models were trained only on datasets of bulk materials, meaning that metal-water interfaces could be a considerable extrapolation.
The only exception is MACE-MH-1, which was also trained on molecular and surface datasets via a multiheaded approach.
For this model we evaluate the 'omatpbe'~\cite{barroso2024open_omat}, 'oc20'~\cite{chanussot2021open_oc20} and 'omol'~\cite{levine2025open_omol} readout heads.
All but one of the foundation models used are trained on the PBE functional, and we apply the D3 correction on top.
The exception is the 'omol' head, which is trained on the $\omega$B97M-VV10 functional.
We also include two models trained on an iteratively (IT) generated dataset with 2 and 4 message passing layers (2l and 4l) respectively. See section on 'Short-Ranged Model Optimisation' for more details.

As stated in the main text, the foundation models generally produce pathologically bad distributions of the total dipole.
Almost all foundation models seem to have especially poor interfacial dipole orientation; there is no cancellation of the near surface dipole compared to AIMD (and the system specific models).  
MATPES-pbe, and the omatpbe and omol heads of MH-1 in particular seem to have a rapid build up of dipole at the copper-water interface.
The deviation in these distributions for some models may in part be due to large amounts of \ce{H2O} dissociation.
The extent of dissociation for each model is shown in Figure~\ref{fig:S:dissociation}, which tracks the percentage of oxygen atoms that are part of a dissociated water species---\ce{O},\ce{OH} and \ce{H3O}---across the MD trajectory.
One sees \ce{OH} and \ce{H3O} species are almost permanently present for the older foundation models, with the MPA model even generating lone oxygen atoms at times. 
Furthermore, for MPA these hydrogen deficient species tend to exist near the interface, perhaps suggesting why the dipole doesn't build up so drastically into the slab center.
In contrast, for the MACE models trained on reference DFT data of this interface, \ce{H2O} molecules remain intact, and dissociation is just transient.
This implies that the selected foundation models seem inappropriate for simulating aqueous interfaces, which is perhaps unsurprising given that this application is out of their training domain, and they are short-ranged models.
We note in the newest foundation model analysed, MH-1, seems less prone to dissociation of water, especially the omol and oc20 heads, which contain some for of molecular information in their training set. This is with the caveat that the omol head is trained upon a different functional.

One interesting deviation from the above results is that the oc20 head of MH-1 looks to give a similar $P_z(z)$ profile to the SR-MACE. 
The oc20 dataset includes surfaces with adsorbates, which may go some may to explaining the better $P_z(z)$ profile, at least at the interface. 
That the other MH-1 heads produce such different results despite using the same base descriptor as the oc20 head is interesting, and raises some questions about the effectiveness of cross learning in this case.

\subsubsection{Total Dipole Autocorrelation Function}

\begin{figure}
    \centering
    \includegraphics[width=1.0\linewidth]{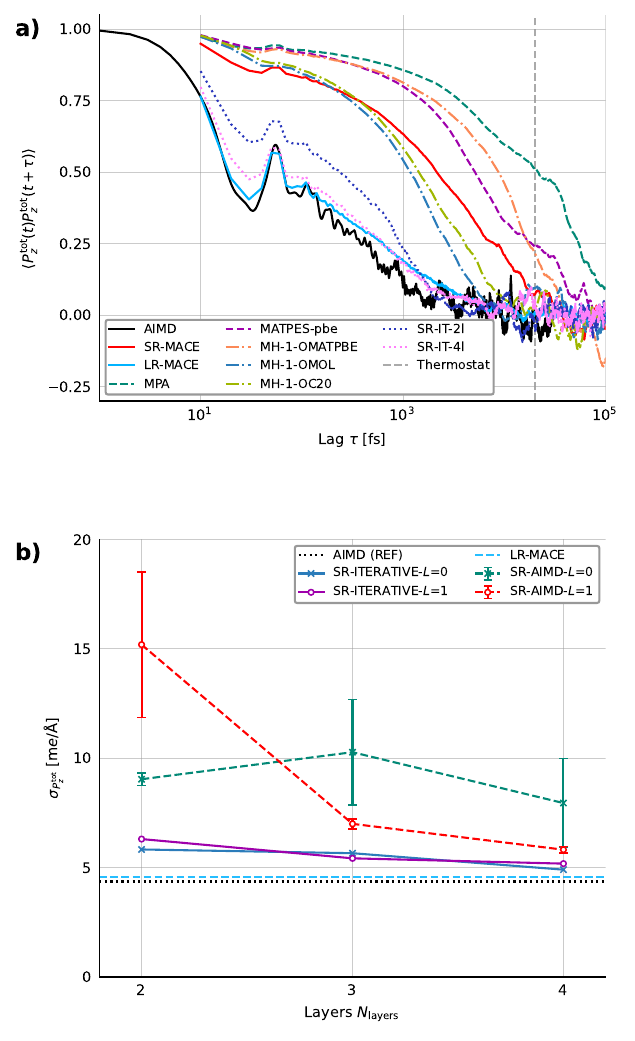}
    \caption{a) Autocorrelation functions of $P^{\text{tot}}_z$ obtained from MD. The vertical grey dashed line shows the timescale of the Langevin thermostat damping $\gamma=0.05\, \text{ps}^{-1}$. Models are the same as those in Figure \ref{fig:S:aimd_profiles_fm}.
    b) Change in standard deviation of $P^{\text{tot}}_z$ with model size and dataset for SR-MLIPs with message symmetry $L$, number of message passing layers $N_{\text{layers}}$. AIMD refers to models trained on a dataset of exclusively AIMD configurations. ITERATIVE refers to models trained on a datase assembled via iterative learning protocol described later in the text. LR-MLIP-MD and AIMD are included for comparison. For the AIMD models, 3 models are trained with different seeds, with the mean and variance being plotted. }
    \label{fig:2:acf_size}
\end{figure}

For an observable $A$ the autocorrelation function $A_{\text{acf}}(\tau)$ at lag $\tau$ is calculated as:
\begin{equation}
    \frac{\sum_i^{T-\tau}
    (A(t_i)-\langle A \rangle )(A(t_i+\tau)-\langle A \rangle )}{(T-\tau)\sigma_A^2}
\end{equation}
where $T$ is total number of frames in the simulation, $t_i$ is the frame number, and $\sigma_A^2$ is the variance of $A$.

The autocorrelation functions for $P_z^{\text{tot}}$ are shown in Figure~\ref{fig:2:acf_size}a.
One sees the LR-MACE matches AIMD closely, whilst the SR-MACEs (including the foundation models) decay much slower. The exception is the four-layer iteratively trained model, which provides very close agreement to the AIMD (though still not as close as the long-ranged model). 
This is discussed in more detail in a later section.
We note that the SR-MLIP autocorrelation functions decay rapidly when $\tau$ is on scale of the Langevin thermostat due to thermal randomisation. 

\subsubsection{Fully Periodic Copper-Water-Copper Example}

Using the 2-layer SR-MACE and LR-MACE trained on AIMD data from Figure~\ref{fig:1:aimd_profiles}, we perform simulations for a 6-layer Cu(111) $5\times6$ with $32\, \text{\AA}$~of water above and no vacuum, corresponding to a ($12.617\times13.112\times 46.311\, \text{\AA}$) unit cell with 181 water molecules.

\begin{figure*}
    \centering
    \includegraphics[width=1\linewidth]{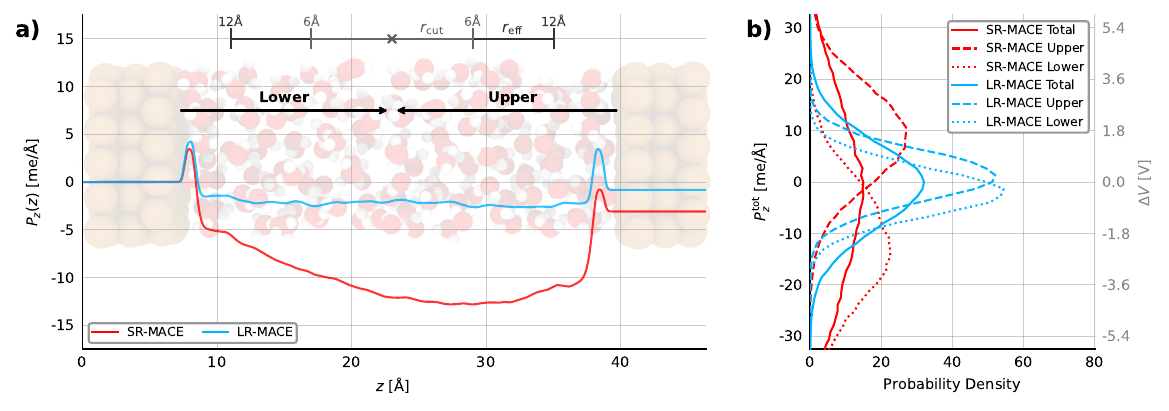}
    \caption{a) The integrated mean dipole per unit area of the water layer $P_z(z)$ across the slab length of a closed copper-water system during MD simulations. 
    The simulation setup is shown in the background. 
    Lower and Upper arrows represent the direction of dipoles from each copper-water interface.
    b) $P^{\text{tot}}_z$ distribution obtained from MD. 
    Histograms of total dipole, and dipole from lower or upper surface up to half way in the cell are shown for each model. Sign of dipole relative to pointing in positive $z$ direction.}
    \label{fig:S:closed_dist}
\end{figure*}

The dipole distributions are shown in Figure \ref{fig:S:closed_dist}, in which one sees a bowing in the $P_z(z)$ profile of the short-ranged model.
This bowing occurs because each interface strongly orders the water around it opposite to the other interface, leading to symmetry around the centre of the water region.
This means the residual ordering artefact that occurs in short-ranged models, which is usually linear across the slab, goes from positive average alignment near one interface, to zero in the centre, to negative at the other.
Integrating this dipole across the slab length leads to a approximately quadratic profile.
We note the profile is not symmetric, likely due to limited sampling time.
Based on this idea of strong interfacial ordering, we also define a water-layer dipole starting from either surface and going up to the slab centre.
We find the distribution of these dipoles, and the total dipole of all water molecules, is broader for the SR-MACE verses the LR-MACE (Figure~\ref{fig:1:aimd_profiles}b), as in the closed system case.

\begin{figure*}
    \centering
    \includegraphics[width=1\linewidth]{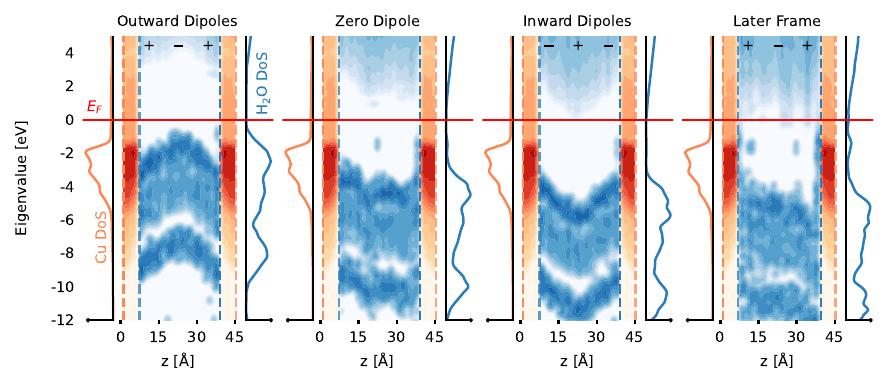}
    \caption{The electronic structure with 0 applied field for a closed $5\times6$ 4 layer Cu(111) slab with $\approx30\, \text{\AA}$ of water above with representative dipoles from each surface, taken from an MD simulation ran using the SR-MACE described in the text. Centrally in each plot is shown the PDoS versus $z$ for copper (orange) and water (blue). 
    Dashed lines represent the highest and lowest atom in each region. The total DoS for the water and copper regions are shown either side. Direction of dipoles indicated by $+,-$ signs. The first three panels are taken from within the first $200\, \text{ps}$ of the simulation. The 'later frame' was taken after $\approx900\, \text{ps}$.
    Exact $P_z$ from each surface to slab centre in m$e$/\AA~are $(-12,13),(-0.2,-0.3),(7,-19),(-34,30)$, in which (a,b) denotes the total dipole of the lower and upper water regions respectively. 
     }
    \label{fig:S:closed_pdos}
\end{figure*}

The PDoS of selected configurations sampled by the SR-MACE are shown in Figure~\ref{fig:S:closed_pdos}.
The PDoS were calculated using the light basis set with a $k$-grid of $3\times3\times1$.
%
The first three panels of Figure~\ref{fig:S:closed_pdos} show three cases relating to the net dipole of the water molecules closest to either the upper or lower interface, as show in Figure~\ref{fig:S:closed_dist}a:
i) where these surface specific dipoles point toward the interface,
ii) where there is $\approx0$ dipole on each interface and
iii) where both dipoles are facing away from the interface. 
Values of these dipole components are given in the caption of Figure \ref{fig:S:closed_pdos}.
We again observe a mechanism for metallization to occur.
This time, the metallization occurs by the water VBM/CBM at the centre of the slab touching the Fermi level, with the bands bending in a similar fashion to the profile of $\langle P_z(z)\rangle$.
This can be rationalised by noting the potential difference is proportional to the integrated dipole up to that point in the slab.
We also note that frames taken later in the MD trajectory have very distorted electronic structure, with an example frame shown in the final panel of Figure~\ref{fig:S:closed_pdos}.

\subsection{Pure Water System}

\subsubsection{MACE Model and Classical Force Field Details}

\begin{figure*}
    \centering
    \includegraphics[width=1\linewidth]{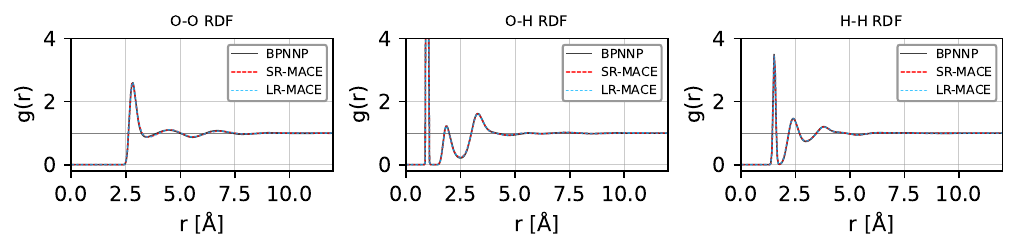}
    \caption{Water radial distribution functions $g(r)$ for cubic $30\, \text{\AA}$ periodic water box obtained from revPBE0-D3 water models. 
    LR-MACE refers to LR-MACE-a in Table~\ref{tab:S:revpbe0errors}.
    Dark grey horizontal line represents a normalised density of 1.}
    \label{fig:S:bulk_water_rdf}
\end{figure*}

For the liquid water simulations, the DFT training set was taken from \cite{schran_committee_2020} and includes various bulk water, water slab, and ice configurations evaluated using the revPBE0-D3 functional \cite{perdew1996generalized_pbe,zhang1998comment_revpbe,adamo1999toward,grimme_dispersion-corrected_2016}. 
We trained several different MACE models on this data, detailed in Table~\ref{tab:S:revpbe0errors}. 
For our water slab experiments we use those labelled SR-MACE and LR-MACE-a models.
We further validated these models by comparing the radial distribution functions (RDFs) for bulk water to the validated BPNNP model from Ref.~\citenum{schran_committee_2020}, which can be seen in Figure \ref{fig:S:bulk_water_rdf}.
We simulated a $30\times30\times30\, \text{\AA}$ cubic water box, details of which can be found in the later section containing 'additional results for water slabs'.
We see that all models trained on this dataset produce near identical RDFs. 

For the comparison against a classical force field, we used a flexible version of the SPC/E water model \cite{berendsen1987missing_spce}.
The bonds and angles are described as harmonic springs, the interactions between all atoms are governed by their charges and Lennard-Jones potentials, with parameters detailed in Table~\ref{tab:FF}.
We note that when calculating $P_z^{\text{tot}}$ for the FF model, we use $q=0.8476$ in equation~\ref{eq:dipole_calc_main}, as this should give us the exact value of $P_z^{\text{tot}}$ for this potential. 
%
Training data was generated from NPT trajectories of water in bulk and small slabs of approximately 60 water molecules each using the FF. 
Bulk and aqueous NaCl, with concentrations from 1 to 8$\, \textsc{m}$ were also included, predominantly for applications in a separate project.
In total, the dataset consists of 13,764 structures. About 40\% are slab structures and the remainder are bulk structures. 


For simulations using the FF, the LAMMPS Molecular Dynamics Simulator was used \cite{LAMMPS}. 
All simulations were done at $300\, \text{K}$ in the NVT ensemble, using a timestep of $0.5\, \text{fs}$.
The simulations with the classical FF were run for 10$\, \text{ns}$ each.
The simulations with the FF-trained MACE models were at least 800$\, $ps long for the largest system (576 atoms, $12.905 \times 12.905 \times 238.715\, \text{\AA}$ cell), 1$\, $ns for the medium sized slab (384 atoms, $12.905 \times 12.905 \times 225.810\, $\AA{} cell), and 2$\, $ns for the smallest system (384 atoms, $16.259 \times 16.259 \times 216.259\, $\AA{} cell). The simulations with the DFT-trained LR-MACE models were ran for $2.8\, \text{ns}$ for the 384 atom systems and $1.8\, \text{ns}$ for the 576 atom system. The simulations with the DFT-trained SR-MACE models were all ran for $3.9\, \text{ns}$.
Configurations were analysed every $250\, \text{fs}$. 

\begin{table*}
    \centering
    \caption{Parameters of the classical force field used as a reference.}
    \begin{tabular}{|c|c||cc|}
        \hline
        Parameter & Value & Lennard-Jones-Parameter & Value \\ \hline
        $r_\text{OH}^\text{eq}$ & $1.012\, \text{\AA}$ & $\sigma_\text{OO} $ & $ 3.165492\, \text{\AA}$ \\
        $k_r$ & $1059.162\, \text{kcal}/\text{mol}/\text{\AA}^2$ & $\epsilon_\text{OO} $ & $0.1554253\, \text{kcal}/\text{mol}$ \\
        $\angle_\text{HOH}^\text{eq} $ & $ 113.246^\circ$ & $\sigma_\text{NaNa} $ & $ 2.583\, \text{\AA}$ \\
        $k_\angle $ & $75.90\, \text{kcal}/\text{mol}/\text{rad}^2$ & $\epsilon_\text{NaNa} $ & $ 0.1 \, \text{kcal}/\text{mol}$ \\
        $q_\text{O} $ & $ -0.8476\, e$ & $\sigma_\text{ClCl} $ & $ 4.4\, \text{\AA}$ \\
        $q_\text{H} $ & $+0.4238\, e$ & $\epsilon_\text{ClCl} $ & $ 0.1 \, \text{kcal}/\text{mol}$\\ \hline
    \end{tabular}
    \label{tab:FF}
\end{table*} 

\subsubsection{Additional results for water slabs}
\label{sec:SI-water-slabs}

\begin{figure*}
    \centering
    \includegraphics[width=1\linewidth]{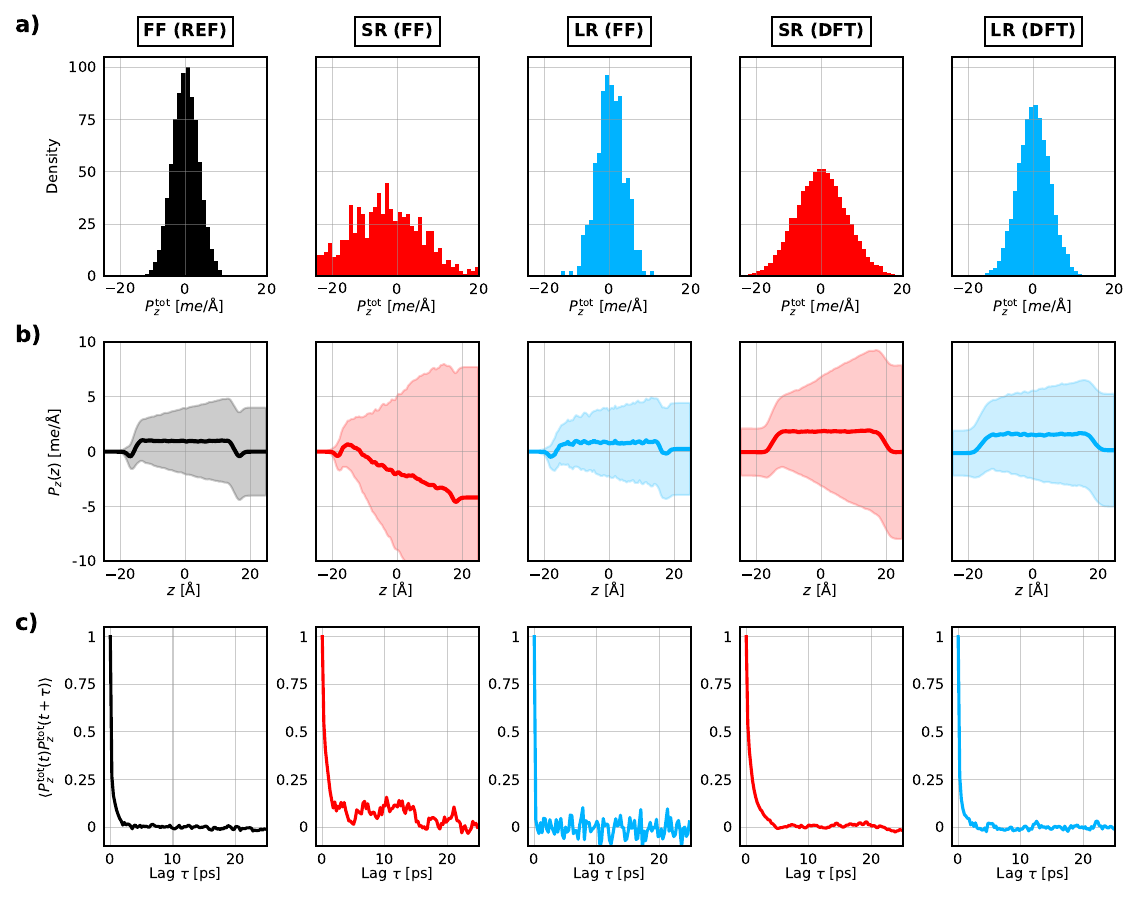}
    \caption{
    Distributions of dipole obtained from MD simulations of $38\, \text{\AA}$ thick water slabs using LR- and SR-MACE models trained on Force Field (FF) data and DFT data. Reference FF simulation is also shown.
    a) Distribution of $P^{\text{tot}}_z$ as histogram.
    b) the cumulative dipole $P_z(z)$ across the length of water slab.
    Shaded regions in these plots represent the standard deviation of $P_z(z)$.
    c) Autocorrelation function of $P_z^\text{tot}$}
    \label{fig:S:water_slab_hist}
\end{figure*}

Figure~\ref{fig:S:water_slab_hist}a shows the histogram of $P_z^{\text{tot}}$ for each model in Figure \ref{fig:2:water_fig}a, specifically for a $38\, \text{\AA}$ thick slab.
The LR-MACE better resembles the FF model distribution, and produces a narrower $P_z^\text{tot}$ relative to the SR-MACE for both the DFT and FF scenarios. 
Figure~\ref{fig:S:water_slab_hist}b shows the average $P_z(z)$ profiles.
For the FF data, the LR-MACE faithfully reproduces the FF reference distributions.
The SR-MACE distribution on the other hand produces a large tilting in it's profile, which may be due to finite sampling statistics and the slow decorrelation time shown. 
Indeed, looking at Figure~\ref{fig:S:water_slab_hist}c we again see the SR-MACE models have slower decorrelation times for the total dipole.
In the DFT case, both the average $P_z(z)$ profiles are quite similar.
The shaded regions of the $P_z(z)$ profiles represent the standard deviation in $P_z(z)$.
We see the standard deviation grows across the slab length, with the rate of increase being notably larger for the SR models.
This reinforces our finding that the variance increases linearly with slab width, and that metallization becomes more prevalent for SR-MLIPs when simulating larger systems.

\begin{figure*}
    \centering
    \includegraphics[width=1\linewidth]{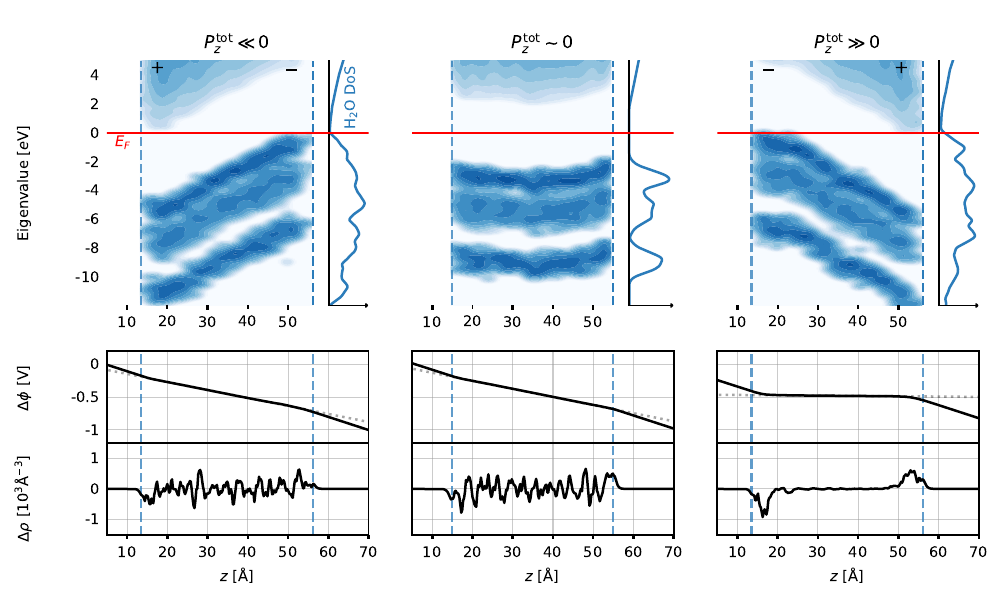}
    \caption{a) The electronic structure with 0 applied field for a ($12.905 \times 12.905 \times 238.7\, \text{\AA}$) cell for $\approx38\, \text{\AA}$ thick water slab taken from an MD simulation ran using the SR model described in the text. Centrally in each plot is shown the PDoS versus $z$ for water. 
    Dashed lines represent the highest and lowest atom in each region. The total DoS for the water is shown on the right. Direction of dipole indicated by $+,-$ signs.
    b) Difference in potential $\phi$ and electron density $\rho$ with $z$ between applied fields of 0 and $0.02\, \text{ev/\AA}$ in the $z$ direction. The dotted grey line shows the gradient of $\Delta \phi$ in the water region.
     }
    \label{fig:S:water_slab_pdos}
\end{figure*}

The PDoS for three slabs generated from the SR-MACE MD are shown in Figure \ref{fig:S:water_slab_pdos}. Again, we chose configurations with a large total dipole to illustrate the band bending effect. Specifically, the three water configurations in the figure have $P_z^\text{tot}$ equal to $-25,0.4$ and $32\, \text{m}e/\text{\AA}$ respectively. 
As mentioned in the main text, we see bending across the entire slab length, meaning that metallization occurs by electrons transferring from one end of the slab into the other.
This differers slightly from the metal-water case as electrons transfer from water-to-water, not water-to-metal, or rephrased from insulator to insulator as opposed to insulator to conductor.
In this case, we only see metallization for $P_z^{\text{tot}} \gg 0$. 
However, the slabs are symmetric around a central mirror plane so we expect this is due to the small statistics.
Also, as the variance in $P_z^\text{tot}$ increases linearly with slab thickness, we anticipate a substantial increase in configurations suffering metallization would occur for a small increase in slab length beyond those simulated here.

\subsubsection{Additional results for bulk water}

\begin{table}[h]
        \centering
        \caption{Validation errors for the models trained on the revPBE0-D3 water data used throughout letter to simulate pure water systems. 
        Energy errors are given in m$e$V/atom and force errors are given in m$e$V/\AA. 
        All MACE models listed here use 2 message passing layers.}
        \begin{tabular}{|cc|cc|cc|}
            \hline
            Model & $r_{\text{cut}}$ & $L$ & $K_{\text{channels}}$ & RMSE($E$) & RMSE($F$) \\
            \hline
            SR-MACE & 6.0 & 1 & 128 & 0.3 & 27.4 \\
            LR-MACE-a & 6.0 & 1 & 128 & 0.18 &  22.89 \\
            LR-MACE-b & 5.0 & 1 & 128 & 0.23  &  22.57 \\
            LR-MACE-c & 5.0 & 0 & 64 & 0.27 & 24.44 \\
            \hline
            BPNNP\cite{schran_committee_2020} & 6.35 & - & -  & 1.8 & 88 \\
            \hline
        \end{tabular}
        \label{tab:S:revpbe0errors}
    \end{table}

    \begin{table}
        \centering
        \caption{Cell size and corresponding number of atoms for the periodic water simulations.
        Cell lengths given to 2 significant figures.}
        \begin{tabular}{|c|cc|}
            \hline
            Cell Size [\AA] & Number Atoms & Number of Molecules\\
            \hline
            $14\times14\times15$ & 315 & 105\\
            $14\times14\times30$ & 630 & 210\\
            $14\times14\times45$ & 945 & 315\\
            $14\times14\times60$ & 1260 & 420\\
            $30\times30\times30$ & 2760 & 840\\
            $45\times45\times45$ & 9315 & 3105\\
            $54\times54\times54$ & 16200 & 5400\\
            \hline
        \end{tabular}
        \label{tab:S:num_water_in_boxes}
    \end{table}

    \begin{table}
        \centering
        \caption{Simulation details for periodic water cell simulations. Models refer to those described in Table \ref{tab:S:revpbe0errors}.}
        \begin{tabular}{|c|c|cc|}
            \hline
            Model & Cell Size [\AA] & Number of Runs & Time [ps] \\
            \hline
            BPNNP & $14\times14\times15$ & 1 & 1000\\
            BPNNP & $14\times14\times30$ & 1 & 1000\\
            BPNNP & $14\times14\times45$ & 1 & 1000\\
            BPNNP & $14\times14\times60$ & 1 & 1000\\
            BPNNP & $30\times30\times30$ & 1 & 1500\\
            BPNNP & $45\times45\times45$ & 1 & 105\\
            \hline
            SR-MACE & $14\times14\times15$ & 1 & 205\\
            SR-MACE & $14\times14\times30$ & 1 & 205\\
            SR-MACE & $14\times14\times45$ & 1 & 205\\
            SR-MACE & $14\times14\times60$ & 1 & 200\\
            SR-MACE & $30\times30\times30$ & 1 & 500\\
            SR-MACE & $45\times45\times45$ & 10 & 120\\
            SR-MACE & $54\times54\times54$ & 10 & 120 \\
            \hline
            LR-MACE-a & $14\times14\times15$ & 1 & 205\\
            LR-MACE-a & $14\times14\times30$ & 1 & 105\\
            LR-MACE-a & $14\times14\times45$ & 1 & 75\\
            LR-MACE-a & $14\times14\times60$ & 1 & 1000\\
            LR-MACE-a & $30\times30\times30$ & 1 & 500\\
            LR-MACE-b & $45\times45\times45$ & 10 & 120\\
            LR-MACE-c & $54\times54\times54$ & 10 & 120 \\
            \hline  
        \end{tabular}
        \label{tab:S:simulation_specs_for_bulkwater}
    \end{table}

\begin{figure}
    \centering
    \includegraphics[width=1\linewidth]{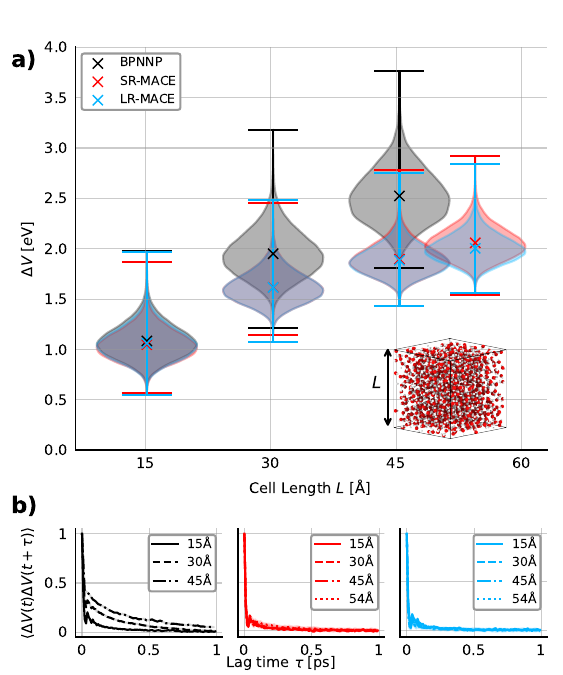}
    \caption{a) Violin plots of the cell potential difference $\Delta V$ using different MLIPs for a cubic bulk water box with cell length $L$. 
    b) Autocorrelation plots of $\Delta V$ for each model and cell size.}
    \label{fig:S:cubic_water_V}
\end{figure}

\begin{figure}
    \centering
    \includegraphics[width=1\linewidth]{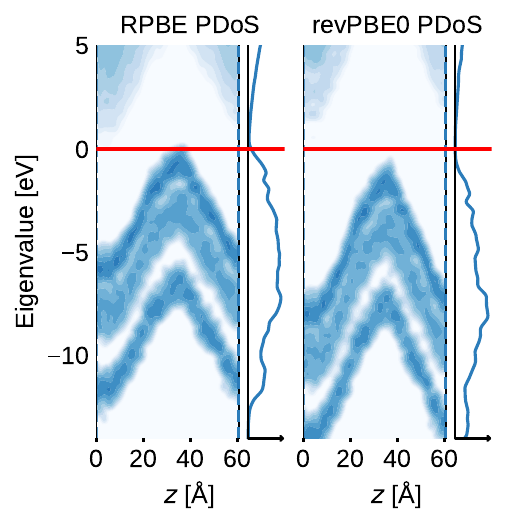}
    \caption{Projected density of states of sampled bulk water configuration with highest cell potential difference from Figure \ref{fig:2:water_fig}c in main text. PDoS were calculated using the RPBE and revPBE0 functionals.
    The combined density of states is shown to the right.}
    \label{fig:S:rod_water_pdos}
\end{figure}

The estimated potential in the periodic cells is obtained by first placing spherical Gaussian charges of $-0.5562e$ on O atoms and $+0.2781e$ on H atoms.
The potential is then interpolated as a Coulomb potential from these smooth charge distributions.
The potential difference is obtained by subtracting the lowest potential on any atom in the simulation cell from the highest potential on any atom.
While this metric is not strictly the same as the DFT potential, it is a good proxy to reveal qualitative trends in band bending.

In Figure~\ref{fig:2:water_fig}c/d of the main text, we simulated elongated periodic cells of dimensions ($a\times a\times L_aN$), with $N=[1,2,3,4]$, $a=14.412\, \text{\AA}$ and $L_a=15.133\, \text{\AA}$.
In Figure \ref{fig:S:cubic_water_V}, we further evaluated the potential difference for cubic cells of size ($L_aN \times L_aN\times NL_a$), with $N=[1,2,3,3.6]$.
Enough water molecules were added to each cell to achieve a density of $\rho=1.00\, \text{g cm}^{-3}$. 
The number of molecules in each cell is given in Table~\ref{tab:S:num_water_in_boxes}.

For the most part, the MACE models used are the same as for the DFT (revPBE0-D3) water slab simulations, with the BPNNP taken from the same reference and trained on the same data.
In all simulations a timestep of $0.5\, \text{fs}$ in combination with hydrogen masses for hydrogen atoms was used, and configurations were saved every $10\, \text{fs}$.
BPNNP simulations were performed using CP2K \cite{kuhne2020CP2K} using the CSVR thermostat \cite{bussi2007canonical} with a time constant of $30\, \text{fs}$. 
The SR-MACE and LR-MACE were performed by interfacing with the Atomic Simulation Environment, using a Langevin Thermostat MD with damping parameter $\gamma=0.05\, \text{ps}^{-1}$. 

Due to memory constraints arising from the increased computational overhead for the LR-MACE, the model size was reduced for the simulations of the cubic cells with $N=[3,3.6]$.
For $N=3$, we used a $r_{\text{cut}}=5.0\, \text{\AA}$, labelled LR-MACE-b.
For $N=3.6$, we used $L=0$, $K_{\text{channels}}=64$ and $r_{\text{cut}}=5.0\, \text{\AA}$, labelled LR-MACE-c. 
The reduction in the size of the short-ranged part of the model is justified on the grounds that in liquid water, the coulomb interaction is the most important non-local interaction.
As we are now modelling this explicitly, our local descriptor can afford to consider a smaller radius around each atom and be less expressive.
These smaller LR-models produce similarly small errors as the larger one, as shown in Table~\ref{tab:S:revpbe0errors}. 
We perform 10 independent MD runs for these systems to enable estimation of uncertainties.

The results of the MD simulations for the cubic box are shown in Figure \ref{fig:S:cubic_water_V}, which shows the same trends as the elongated cell, albeit the rate of divergence of the variance is found to be slightly slower. 
This shows that while geometry plays some role in the exact change of the distribution, for large enough cells the large fluctuations in cell potential difference in SR-MLIPs seems to be a universal phenomenon.
This relates to previous findings in the literature that the molecular dipole density in reciprocal space diverges for small reciprocal space wavelengths, corresponding to large real space distances \cite{cox_dielectric_2020}.

We evaluated the PDoS for the elongated water cell with the largest potential difference using the revPBE0 and RPBE \cite{hammer1999improved_rpbe} functionals, shown in Figure \ref{fig:S:rod_water_pdos}.
This corresponds to $L_z\approx60\, \text{\AA}$, and was generated by the BPNNP.
The band gap closes for the RPBE calculation, and has nearly closed for the revPBE0 functional. 
Being a hybrid functional, the revPBE0 functional is expected to have a larger band gap on account of the reduced self-interaction error.
The estimated potential difference in this cell was $\approx 6.5\, e\text{V}$ (see Figure~\ref{fig:2:water_fig}c), which, looking at the revPBE0 PDoS, corresponds to the degree of band bending observed in the PDoS.
Metallization occurs for the RPBE functional before the bands can bend to this extent, highlighting another case where our potential proxy based on localised atomic charges breaks down.

\subsection{Short-Ranged Model Optimisation}
\label{iterative-section}

\begin{table*}
    \centering
    \caption{Root mean square error of energies $E$ and forces $F$ on validation set of models used to compare effect of model hyperparameters and dataset on dipole distribution in Figure~\ref{fig:2:acf_size}b. RMSE($E$) given in m$e$V/atom and RMSE($F$) in m$e$V/\AA. The LR-MACE is trained on AIMD data, and the model used previously in the text (see Table~\ref{tab:S:aimd_models_cuw}.
    }
    \begin{tabular}{|l|ccc|cc|}
    \hline
        Model & Version & $N_{\text{Layers}}$ & $L$  & RMSE($E$) & RMSE($F$) \\
\hline
AIMD & 1 & 2 & 0 & 0.2 & 49.5 \\
AIMD & 1 & 2 & 1 & 0.1 & 48.7 \\
AIMD & 1 & 3 & 0 & 0.1 & 47.9 \\
AIMD & 1 & 3 & 1 & 0.1 & 47.6 \\
AIMD & 1 & 4 & 0 & 0.1 & 47.8 \\
AIMD & 1 & 4 & 1 & 0.1 & 47.4 \\
\hline
AIMD & 2 & 2 & 0 & 0.2 & 49.5 \\
AIMD & 2 & 2 & 1 & 0.1 & 48.3 \\
AIMD & 2 & 3 & 0 & 0.1 & 48.1 \\
AIMD & 2 & 3 & 1 & 0.2 & 47.6 \\
AIMD & 2 & 4 & 0 & 0.1 & 48.7 \\
AIMD & 2 & 4 & 1 & 0.2 & 47.4 \\
\hline
AIMD & 3 & 2 & 0 & 0.1 & 49.6 \\
AIMD & 3 & 2 & 1 & 0.1 & 49.1 \\
AIMD & 3 & 3 & 0 & 0.2 & 48.0 \\
AIMD & 3 & 3 & 1 & 0.1 & 47.6 \\
AIMD & 3 & 4 & 0 & 0.1 & 48.7 \\
AIMD & 3 & 4 & 1 & 0.2 & 47.4 \\
\hline
ITERATIVE & 1 & 4 & 0 & 0.3 & 47.8 \\
ITERATIVE & 1 & 4 & 1 & 0.7 & 47.9 \\
ITERATIVE & 1 & 3 & 0 & 0.3 & 47.7 \\
ITERATIVE & 1 & 2 & 1 & 0.4 & 48.7 \\
ITERATIVE & 1 & 2 & 0 & 0.4 & 49.1 \\
ITERATIVE & 1 & 3 & 1 & 0.5 & 48.0 \\
\hline
LR-MACE & 1 & 2 & 1 & 0.14 & 46.23 \\
\hline
    \end{tabular}
    
    \label{tab:iterate_model_table}
\end{table*}

This section relates to training approaches used to minimise the extent of false solvent metallization in short-ranged models.
There are two common approaches to improving short-ranged MLIPs which we consider: i) increasing the model size and expressivity, and ii) increasing the data set size and diversity, often through some sort of active learning or iterative procedure.

To alter the models expressivity we change two hyperparameters:
We increase the number of message passing layers $N_{\text{layers}}$ to linearly increase the effective radial cutoff, specifically trying $N_{\text{layers}}=2,3,4$.
One might hope that the model can then capture more of the long-ranged component of the electrostatic interaction.
We extend $N_{\text{layers}}$ as opposed to $r_\text{cut}$ due to better computational scaling; cost scales approximately linearly with number of layers ($\mathcal{O}(N_{\text{layers}})$) versus approximately cubicly with cutoff $\mathcal{O}(r_{\text{cut}}^3)$.
We simultaneously assess the impact of using an invariant model $(L=0)$ versus an equivariant $(L=1)$ model. 
Equivariant features are able to capture relative orientations of atoms molecules outside the local cutoff.
As the metallization arises from over-alignment of molecules, one may assume it is easier to avoid for equivariant models.

To evaluate the effect of dataset, we expand it via iterative learning.
This is to disentangle problems arising from model architecture from problems arising from the MLIP incorrectly extrapolating outside its training set. 
Our iterative training protocol is as follows:
We first train a model only on configurations sampled during AIMD, which we then use to sample configurations with during an MD run using the same simulations settings as the models in Figures~\ref{fig:1:aimd_profiles} and \ref{fig:S:aimd_profiles_fm}.
Configurations with dipoles not present in the AIMD data and higher levels of dissociation of water are targetted to expand the sampled configuration space.
Particularly unphysical looking configurations, such as those containing isolated O atoms in the water, were discarded.
DFT was then performed on each sampled frame using the same DFT settings as for the AIMD, and models are retrained on this dataset.
Additionally, configurations from the LR-MLIP MD run were added as a proxy for a long AIMD run.
Three iterations of adding configurations from SR-MLIP-MD were performed.
The number of configurations added per iteration are summarised in Table~\ref{tab:S:iterative_training}. 
The final dataset, labelled 'ITERATIVE', was used for training a new set of models.
 
A model was trained for each combination of dataset, $N_{\text{layers}}$ and $L$.
For the AIMD dataset, three models were trained using a different seed to measure the variation in performance.
Each simulation was ran for $205\, \text{ps}$, with the same simulation settings as for Figure \ref{fig:1:aimd_profiles}.
Frames were analysed every $10\, \text{fs}$, and the first $5\, \text{ps}$ discarded.

The errors for each model are shown in Table~\ref{tab:iterate_model_table}.
One can see that all models produce similar errors in the forces, regardless of architecture and hyperparameters. 
Notably the iterative dataset produces slightly higher energy errors, which may be caused by the model having a less homogeneous dataset with more extreme structures.
We note for $(L=1, N_{\text{layers}}=2)$, the large mean and variance in $\sigma_{P^{\text{tot}}_z}$ arises from instability in one MD run.

The standard deviation of $P_z^\text{tot}$ during MD for each model is shown in Figure~\ref{fig:2:acf_size}b. 
Generally, for equivariant models ($L=1$) one sees $\sigma_{P_z^{\text{tot}}}$ systematically improves with $N_{\text{layers}}$.
Such a trend is less clear for the $L=0$ models, which points to our hypothesis that equivariant models should be able to better capture the orientational arrangements that lead to metallization.

Regarding the dataset, we see models trained on the 'ITERATIVE' dataset consistently provide better performance than models trained exclusively on the AIMD data, regardless of model hyperparameters. This is likely because the model is not extrapolating outside of its dataset when it encounters an extreme value of $P_z^\text{{tot}}$.
Combining all factors, one can produce an MLIP approaching the AIMD distribution for the copper-water interface.
We note for $N_{\text{layers}}=4$ the invariant and equivariant models give very similar performance.
We also note for the 4 layer model, which has a receptive field of $24\, \text{\AA}$, nearly all water molecules in the $\approx30\, \text{\AA}$ water layer should be in this receptive field.
Finally, we see that the LR-MACE still produces a closer match to the AIMD distribution, despite being trained on less diverse AIMD data, and having $N_{\text{layers}}=2$. 
This shows that modelling long-ranged interactions with a correct functional form is better than trying to capture most of the shorter-ranged component of them correctly.
We again would like to emphasise that this large variation in performance between models comes despite all models having very similar energy and force errors.

\begin{table}
    \centering
    \caption{Number of configurations added per stage of iterative learning whilst training iterative SR-MACE model. Gen3 dataset is the one used to perform production runs. LR refers to configurations sampled from LR-MACE-MD to augment dataset.}
    \begin{tabular}{|c|c|c|}
    \hline
        Iteration & Configurations added & Total configurations\\
    \hline
        Gen0 & +243 & 243 \\
        LR   & +547 & 790 \\
        Gen1 & +95  & 885 \\ 
        Gen2 & +227 & 1112 \\
        Gen3 & +210 & 1322 \\
    \hline
    \end{tabular}
    
    \label{tab:S:iterative_training}
\end{table}

\begin{figure*}
    \centering
    \includegraphics[width=1\linewidth]{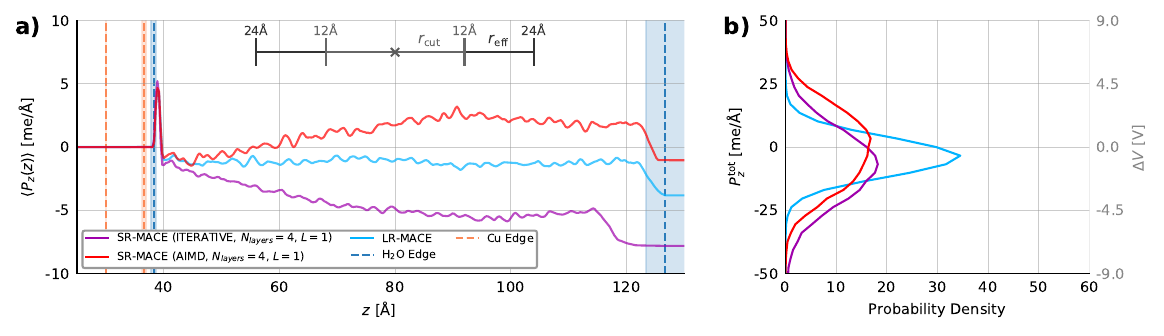}
    \caption{a) The integrated mean dipole per unit area of the water layer $P_z(z)$ across the slab length of a copper-water systems during MD simulations, with water layer $\approx90\, \text{\AA}$ long.
    Short-ranged model radial $r_\text{cut}$ and effective $r_\text{eff}$ cutoff are indicated relative to a point marked by 'X'. 
    Range of the edges of the copper and water regions indicated by blue and orange shaded regions, with dashed line representing mean across simulation.
    b) $P^{\text{tot}}_z$ distribution obtained from MD.
    \label{fig:S:A90}
    }
\end{figure*}

We mention in the main text that the best performing models do not generalise to larger systems.
We assessed this using the same $5\times6$ 4-layer copper (111) surface as before, but then extending the water layer to be $\approx90\, \text{\AA}$ (497 water molecules). We used a cell size of $12.62\times13.11\times154.82\, \text{\AA}$, and the same simulation settings as previous copper water interface simulations.
We simulated this system using the LR-MACE trained on copper-water interfaces, and two $N_{\text{layers}}=4,L=1$ SR-MACE models trained on the purely AIMD and iteratively trained datasets respectively (see Table~\ref{tab:iterate_model_table}). 

The results of the simulations are shown in Figure~\ref{fig:S:A90}. We see, like in Figures~\ref{fig:1:aimd_profiles} and \ref{fig:S:aimd_profiles_fm}, that a gradient re-emerges in the SR-MACE $P_z(z)$ profiles, and that their variance is wider relative to the LR-MACE.
This is despite all models seemingly producing similar performance for the smaller slab systems.
This is also linked to the fact that $1/r$ is a conditionally convergent sum. Going from a smaller to a larger system adds more terms to this sum, impacting our ability to extrapolate.
Overall, the main takeaway from this analysis is that expressive enough short-ranged MLIPs can produce acceptable performance if most atoms are inside the effective cutoff of all other atoms.
These models will fail however when pushed to a larger system size.
For best performance, transferability and correct physicality, one should use a LR-MLIP.

\subsection{Insulating \ce{TiO2}-Water Interface}

\begin{figure*}
    \centering
    \includegraphics[width=1\linewidth]{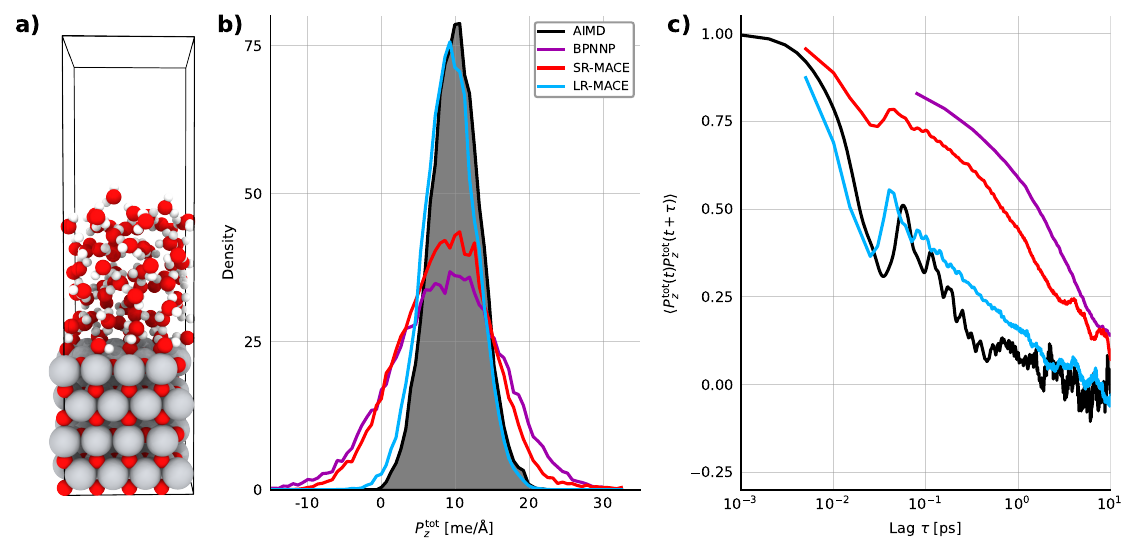}
    \caption{a) Simulation setup of \ce{TiO2} rutile (110) surface slab with water above.
    b) MD distribution of $z$ component of proxy dipole per unit area $P^{\text{tot}}_z$ using difference potentials.
    c) Autocorrelation function of $P^{\text{tot}}_z$.}
    \label{fig:S:tio2}
\end{figure*}

Our analysis for the \ce{TiO2}-water interface is based on BPNNP models and data reported in Ref.~\citenum{schran_machine_2021}.
The \ce{TiO2} rutile(110) was described by a system consisting of 80 water molecules on four O-Ti-O trilayers---a $4\times4\times4$ slab---in a periodic box of dimensions $(11.84\times 12.99\times42.00\, \text{\AA}$). An illustration of the simulation setup can be found in Figure~\ref{fig:S:tio2}a.
The optB88-vdW \cite{klimevs2009chemical_optB88} functional was used in combination with GTH pseudopotential, a $400\, \text{Ry}$ plane wave cutoff and the DZVP molecularly optimised basis set for all elements.
A $1.0\, \text{fs}$ timestep in combination with deuterium masses for hydrogen atoms was used and the atoms of the lowest O-Ti-O trilayer not in contact with water were kept fixed. 
Further computational details can be found in Ref.~\citenum{schran_machine_2021}.
The same dataset was used to train a SR-MACE and LR-MACE. The error's of these models can be found in Table~\ref{tab:S:tio2}.
The SR-MACE and LR-MACE MD were performed with the ASE package using a Langevin thermostat with $\gamma=0.05\, \text{ps}^{-1}$. 

The results of our simulations can be found in Figure~\ref{fig:S:tio2}.
We see similar trends as for the copper water interface in terms of the variance of $P_z^{\text{tot}}$ (panel b) and the autocorrelation function (panel c), showing our findings generalise to insulating interfaces, and also between SR-MLIP architectures.
In particular, both BPNNP and SR-MACE produce dipole distributions that are too wide compared to the AIMD reference, while LR-MACE matches the distribution very well.
Similarly, correlation times are significantly longer for the BPNNP and SR-MACE, while LR-MACE and AIMD feature quickly decorrelating total dipoles.

\begin{table}[h]
    \centering
    \caption{Errors of models on forces and energies used for \ce{TiO2}-water interface, trained on AIMD data.}
    \begin{tabular}{|ccc|}
    \hline
        \shortstack{Model \\ Name}  & \shortstack{RMSE($E$) \\ {[m$e$V/atom]}} & \shortstack{RMSE($F$) \\ {[m$e$V/\AA]}} \\
    \hline
        LR-MACE & 0.14 & 46.23 \\
        SR-MACE & 0.11 & 30.08 \\
        \hline
    \end{tabular}
    
    \label{tab:S:tio2}
\end{table}

\end{document}